\documentclass[12pt]{article}

\global\arraycolsep=1pt
\usepackage{amssymb}
\usepackage{amsmath}
\usepackage {pstricks}
\usepackage{graphicx,color}
\usepackage{mathrsfs}
\newcommand{\beq}{\begin{equation}}
\newcommand{\eeq}{\end{equation}}
\newcommand{\beqa}{\begin{eqnarray}}
\newcommand{\eeqa}{\end{eqnarray}}
\newcommand{\CR}{\nonumber \\}

\newcommand{\lam}{\lambda}

\newcommand{\ep}{\epsilon_1}
\newcommand{\es}{\epsilon_2}

\newcommand{\unitbox}
{\setlength{\unitlength}{0.5pt}
\begin{picture}(10,10)
\put(0,10){\line(1,0){10}}
\put(0,0){\line(1,0){10}}
\put(0,0){\line(0,1){10}}
\put(10,0){\line(0,1){10}}
\end{picture}}
\numberwithin{equation}{section}
\begin{document}

%
\begin{titlepage}
\begin{flushright}
{YITP-12-14}\\
{March, 2012 }
\end{flushright}
\vspace{0.5cm}
\begin{center}
{\Large \bf Generalized Whittaker states for instanton counting 
with fundamental hypermultiplets}
\vskip1.0cm
{\large Hiroaki Kanno$^\dagger$ and Masato Taki$^\ddagger$}
\vskip 1.0em
{\it %
$^\dagger$Graduate School of Mathematics and KMI\footnote[1]
{Kobayashi-Maskawa Institute for the Origin of Particles and the Universe}, \\
Nagoya University, Nagoya, 464-8602, Japan
\vskip 1.0em
and
\vskip 1.0em
$^\ddagger$
Yukawa Institute for Theoretical Physics,\\ 
Kyoto University, Kyoto 606-8502, Japan
}%
\end{center}
\vskip1.5cm

\begin{abstract}
M-theoretic construction of $ \mathcal{N}=2$ gauge theories implies that
the instanton partition function is expressed  as the scalar product of 
coherent states (Whittaker states) in the Verma module 
of an appropriate two dimensional conformal field theory.
We present the characterizing conditions for such states 
that give the partition function with fundamental hypermultiplets
for $SU(3)$ theory and $SU(2)$ theory with a surface operator. We find 
the states are no longer the coherent states in the strict sense
but we can characterize them in terms of a few annihilation operators of lower levels
combined with the zero mode (Cartan part) of the Virasoro algebra $L_0$ or
the $\mathfrak{sl}(2)$ current algebra $J_0^{0}$. 
\end{abstract}
\end{titlepage}


\renewcommand{\thefootnote}{\arabic{footnote}} \setcounter{footnote}{0}


\section{Introduction}

The M5-brane is one of the fundamental dynamical objects 
in M-theory. Its dynamics is still mysterious, though the low
energy effective dynamics is supposed to be governed  by 
six-dimensinal ${\cal N}=(2,0)$ superconformal field theory.
More precisely we have ADE classification of 
${\cal N}=(2,0)$ theory in six dimensions
and the world-volume theory of $N+1$ M5-branes gives
the theory of $A_N$ type. A crucial problem is that 
such six-dimensional ${\cal N}=(2,0)$ theory does not 
allow any Lagrangian description. But one may consider
a compactification to supersymmetric gauge theories in
lower dimensions by wrapping multiple M5-branes 
on a suitable manifold. For example, the compactification on 
$S^1$ gives five-dimensional ${\cal N}=1$ Yang-Mills theory.

The (twisted)  compactification of the six-dimensional ${\cal N}=(2,0)$ theory 
on a Riemann surface $C$ gives rise to diverse ${\cal N}=2$ theories
in four dimensions \cite{Witten:1997sc, Gaiotto:2009we}. 
Such ${\cal N}=2$ theories are therefore labeled by the Riemann surfaces,
on which we can define conformal field theory (CFT).
It is proposed the instanton
partition function of the ${\cal N}=2$ superconformal 
theory agrees with the conformal block of an appropriate
CFT on $C$ \cite{Alday:2009aq, Wyllard:2009hg}. 
In this correspondence,
the choice of ${\cal N}=2$ theory determines the chiral algebra of CFT,
which controls the conformal block.
Furthermore, for non-conformal (asymptotically free) 
cases it is expected that the partition function is expressed as the scalar
product (or the norm) of appropriate states in the 
Verma module \cite{Gaiotto:2009ma, Marshakov:2009gn}. 
This construction gives us a generalization of the conformal block 
in the sense that it has irregular singularities due
to  ``irregular vertex insertions".
For example, for $SU(2)$ gauge theory which corresponds to 
the Virasoro algebra with generators $L_n$, 
the instanton (Nekrasov) partition function of
the pure Yang-Mills theory \cite{Nekrasov:2002qd} is given by
\beq
Z^{(N_f=0)}_{SU(2)} = \langle G_0
\vert 
G_0
\rangle,
\eeq
where we introduce a state $\vert 
G_0
\rangle$
in the Virasoro Verma module with the conformal weight $\Delta$
and the central charge $c$ by the conditions 
\beq
\label{WVpureYM}
L_1 \vert
G_0
\rangle 
= \Lambda^2 \vert 
G_0
 \rangle, \qquad
L_2 \vert G_0 
\rangle = 0.
\eeq
The state $\vert G_0
\rangle$ is called Gaiotto state and expanded 
in the dynamical scale parameter $\Lambda$ 
of the pure $SU(2)$ Yang-Mills theory.

In this paper we will consider the conditions that characterize
such a state for asymptotically free ${\cal N}=2$ theories with 
fundamental matter hypermultiplets. This problem is largely motivated by the 
fact that the conditions \eqref{WVpureYM} 
have indeed the following M-theoretical origin.
The pure $SU(2)$ Yang-Mills theory is the world-volume theory of two M5-branes 
on a Riemann sphere $\mathbb{CP}^1$ with defects at $z=0,\infty$.
Notice that these two defects correspond to the boundaries of D4-branes terminated on 
two parallel NS5-branes
in the Hanany-Witten construction \cite{Hanany:1996ie}.
The low-energy dynamics can be collected into 
the spectral curve $\langle \det (xdz-\Phi)\rangle=0$ for  a one-form field $\Phi$ on $\mathbb{CP}^1$;
\begin{align}
\label{GpureYM}
x^2\left({dz}\right)^2=
\left(\frac{ \Lambda^2}{z}
+{2u}
+{ \Lambda^2 }{z}\right)
\left(\frac{dz}{z}\right)^2
=:\phi_2(z),
\end{align}
which gives the Seiberg-Witten curve.
This curve is specified
by a quadratic differential $\langle\textrm{Tr}~\Phi^2\rangle=\phi_2(z)$ with poles at
the insertion loci of defects.
The point is that this quadratic differential is translated into 
the energy-momentum tensor $T(z)=T_{zz}(dz)^2$ of CFT as
$\langle G_0 \vert T(z)\vert G_0 \rangle
\sim
\phi_2(z)
$.
By using the mode-expansion $T_{zz}=\sum L_n z^{-2-n}$
and the proposed dictionary $u\sim \Delta$ of \cite{Alday:2009aq},
we can encode the information (\ref{GpureYM}) of the defects into the conditions (\ref{WVpureYM})
for the state which describes the singularity at $z=0,\infty$.

For pure Yang-Mills theories without matter hypermultiplets
this story has been generalized to $SU(N)$ theory
and those with a surface operator \cite{Mironov:2009by}--\cite{Gaiotto:2012}. 
In particular in \cite{Kanno:2011fw} the conditions for the state
whose norm gives the instanton partition function 
are proposed for $SU(N)$ theory with a surface
operator of general type. All these states are
characterized as coherent states of generalized $W$ algebra,  
since they are defined as a simultaneous 
eigenstate of the annihilation operators for the
highest weight state of the Verma module. 
In mathematics such states are often called Whittaker states.
It is a natural question if the characterization of the state 
as a Whittaker state is valid for asymptotically free 
theories with matter hypermultiplets. As far as we know,
this problem is worked out only for $SU(2)$ theory, 
where one can introduce
another Gaiotto state $\vert  G_1, m \rangle$ that satisfies
\beq
L_1 \vert  G_1, m \rangle 
= (\epsilon_{+} -2m) \Lambda  \vert  G_1, m \rangle, \qquad
L_2 \vert  G_1, m \rangle 
= - \Lambda^2 \vert  G_1, m \rangle, \label{Nf1}
\eeq
where $m$ is the mass parameter and 
$\epsilon_{+}:= \ep + \es$ is the sum of the $\Omega$-background 
parameters, or the equivariant parameters of the toric 
action on $\mathbb{C}^2$. Then
the partition function of $N_f =1,2$ theory is
given by
\beq
Z_{SU(2)}^{(N_f=1)} = \langle  G_0 
\vert  G_1, m \rangle,
\qquad
Z_{SU(2)}^{( N_f=2)} = \langle  G_1, m_1
\vert \ G_1, m_2 \rangle,
\eeq
respectively\footnote{ Precisely speaking the second equality is
valid up to a contribution from $U(1)$ theory. }.
These proposals are proved in \cite{Fateev:2009aw,Hadasz:2010xp}.
Unfortunately their proof works only for the Virasoro conformal block.
In this paper we will generalize the definition of
states like \eqref{Nf1} into two directions. One is to make the rank of the
gauge group higher and the other is to introduce
a surface operator. Thus, we consider
$SU(3)$ theory and $SU(2)$ theory with a surface operator. 
For $SU(3)$ theory the chiral algebra on CFT side is $W_3$ algebra with generators 
$L_n$ and $W_n$ \cite{Wyllard:2009hg}, while it is the untwisted affine algebra 
$A_1^{(1)}$ or $\mathfrak{sl}(2)$ current algebra with generators 
$J^{\pm, 0}_n$ for the latter \cite{Alday:2010vg}. 
For the $SU(3)$ theory with $N_f=1$  we find the following conditions;
\beqa
L_1 \vert  G_1, m \rangle 
&=& i \frac{ \Lambda_{SU(3)}^2}{\epsilon_1 \epsilon_2}
\vert  G_1, m \rangle, \label{W3L1} \\
W_1 \vert  G_1, m \rangle
&=& \sqrt{\frac{27}{4 \epsilon_1 \epsilon_2 + 15 \epsilon_{+}^2}}  
\frac{(2 m- \epsilon_{+} ) \Lambda_{SU(3)}^2}{2 \epsilon_1 \epsilon_2}
\vert  G_1, m \rangle, \label{W3W1}
\eeqa
and $\vert  G_1, m \rangle$ is annihilated by $L_{n\geq 2}$ and $W_{n \geq 2}$.
Thus, as in the case of $SU(2)$ theory, 
the state $\vert  G_1, m \rangle$ is still a coherent state of $W_3$ algebra, even if 
we add a fundamental matter. However, 
for $SU(2)$ theory with a surface operator, we obtain
\beqa
\left( J_0^+ + \sqrt{x} J_0^0 \right) \vert G_1,  m \rangle &=& 
\frac{\sqrt{x}}{2 \ep}\left(\ep+ 2m \right)
\vert G_1,  m \rangle,  \label{A1+}  \\
 J_1^0 \vert G_1,  m \rangle &=&  \frac{\sqrt{z}}{2\ep} \vert G_1,  m \rangle,  \label{A10} \\
 J_1^{-} \vert G_1,  m \rangle &=&  \frac{\sqrt{z}}{\ep\sqrt{x}} \vert G_1,  m \rangle, \label{A1-}
\eeqa
where $z, x$ are the parameters of topological expansion of the partition function.
In general a coherent state $\vert \Psi \rangle$ must have
zero eigenvalue for a generator that can be 
expressed as a commutator of two annihilation 
operators. For the Virasoro algebra $L_{n \geq 3}$
are such generators. We note that 
there appears the zero mode $J_0^0$ in \eqref{A1+},
which implies the state $\vert G_1,  m \rangle$ has non vanishing eigenvalue for  $J_1^{0} \sim
[J^{+}_0, J^{-}_1]$. Thus it is not a Whittaker state in the genuine sense,
since the coherent condition for $J^+_0$ involves the zero mode of the current $J^0$.
In this paper we will call it generalized Whittaker state. 
On the Verma module of the $\mathfrak{sl}(2)$ current algebra
the action of  $J_0^0$ can be written as
$J_0^0 = j- \sqrt{x} \frac{\partial}{\partial \sqrt{x}}$, where $j$ is the eigenvalue of the
primary state and the Euler derivative with respect to $\sqrt{x}$ counts
the total $SU(2)$ spin. Due to the dictionary $j =  - \frac{1}{2} + \frac{a}{\ep}$ by \cite{Alday:2010vg}, 
the condition for generalized Whittaker state involves the Coulomb moduli parameter $a$. 
It is quite interesting that the same features also appear
in the definition of the state $\vert G_2, m_1, m_2 \rangle$ 
for the $SU(3)$ theory with $N_f =2$.
The state is an eigenstate of a linear combination of $W_1$ and
the zero mode $L_0$ of the Virasoro subalgebra. The $L_0$ part
introduces the dependence on the Coulomb moduli $a_{1,2}$ of $SU(3)$ theory.
Finally the state has non-vanishing eigenvalues for $W_2$ and $W_3$ 
which are commutators of lower generators.

From the viewpoint of M theory the existence of
such generalized Whittaker states is understood as follows;
the $S^1$ compactification of the six-dimensional ${\cal N}=(2,0)$ theory
gives a five-dimensional Yang-Mills theory on $\mathbb{R}^4 \times 
\mathbb{R}_t$ with the maximal supersymmetry. If we consider 
the large volume limit of the space-like part; $\mathrm{Vol}~(\mathbb{R}^4) >> 1$,
we obtain a supersymmetric quantum mechanics on the instanton moduli space
with the wavefunction $\Psi(t) : \mathbb{R}_t \to \mathcal{H}(\mathcal{M}_{\mathrm{inst}})$.
Let us further assume the time direction is a segment of length $\ell$ and put
boundary conditions at $t=0$ and $t=\ell$ which break half of the supersymmetry.
Then the partition function of our system is
\beq
Z_{\mathrm{SQM}} = \langle \Psi_{\mathrm{f}} \vert e^{-\ell H} \vert \Psi_{\mathrm{i}} \rangle,
\eeq
where $\Psi_{\mathrm{i,f}}$ is in the subspace of BPS states $\mathcal{H}_{\mathrm{BPS}}
\subset  \mathcal{H}(\mathcal{M}_{\mathrm{inst}})$
that correspond to the boundary conditions at $t=0, \ell$. We can identify $Z_{\mathrm{SQM}}$
with the Nekrasov partition function, where $e^{-\ell H}$ gives rise to the parameter of 
instanton expansion by $\Lambda^{2N_c-N_f}=e^{-\ell}$. The point here is that
an appropriate generalized $W$ algebra which is obtained by the Hamiltonian reduction 
of the $\mathfrak{sl}(n)$ current algebra, acts  (at least) 
on the BPS subspace $\mathcal{H}_{\mathrm{BPS}}$
and consequently it can be identified with the Verma module of the generalized 
$W$ algebra\footnote{It is not necessarily irreducible.}.
The Hamiltonian of the quantum mechanics then acts on $\mathcal{H}_{\mathrm{BPS}}$
as the Virasoro zero-mode $L_0$,
so that the instanton expansion is just the level expansion on the CFT side.
The corresponding CFT is obtained by 
taking an opposite limit\footnote{
This ``compactification of $\mathbb{R}^4$" is materialized by introducing $\Omega$-deformation
\cite{Nekrasov:2002qd} which localizes the field configurations to the origin.}
; $\mathrm{Vol}~(\mathbb{R}^4) << 1$,
and we expect we can find the states in the Verma module which correspond to
$\Psi_{\mathrm{i,f}}$. These are our generalized Whittaker states. Note that
the world sheet of this CFT is a cylinder and if $S^1$ shrinks it becomes 
a sphere with two puntures. We thus recover the sphere that appeared above
in connection with $SU(2)$ Seiberg-Witten curve. We expect such a curve
would be related to the spectral curve of the Hitchin system in general. 
It is desirable to understand the conditions derived in this paper
from the viewpoint of the Hitchin system.

This article is organized as follows; 
In section 2, we assume that the state $\vert  G_1, m \rangle$ is
a simultaneous eigenstate of $L_1$ and $W_1$ and
derive the conditions \eqref{W3L1} and \eqref{W3W1}
by comparing the one instanton partition function with $N_f=1,2$ and the level one 
Shapovalov matrix of $W_3$ algebra. Then we do the same
for the state $\vert G_2, m_1, m_2 \rangle$ with two mass
parameters and the partition function with $N_f=3,4$.
It turns out that the level one \lq\lq eigenvalue\rq\rq\ of $W_1$ 
for $\vert G_2, m_1, m_2 \rangle$ depends
on the Coulomb moduli $a_{1,2}$ of $SU(3)$ theory. 
In section 3 we test the level one results in section 2 by computing
the decoupling limit from the superconformal theory with $N_f=6$.
We find that at higher levels the Coulomb moduli dependence 
of the $W_1$-eigenvalue should be promoted to the action of 
the Virasoro zero mode $L_0$ and consequently 
the state $\vert G_2, m_1, m_2 \rangle$ cannot be a $W_1$-eigenstate.
In section 4 we understand our results from the viewpoint of
the Seiberg-Witten curve. We show that the phenomena found 
in section 3 take place in general for $SU(N)$ theory
with $N_f=N-1$. In section 5 we derive the conditions \eqref{A1+} -- \eqref{A1-}
from the decoupling limit of the correspondence of the instanton partition 
function with a surface operator and the affine conformal block of $A_1^{(1)}$ 
current algebra proposed by Alday-Tachikawa. 
Some of the technical details are collected in appendices. 



\section{One instanton computation of $SU(3)$ theory}

The AGT relation \cite{Alday:2009aq} between four-dimensional $\mathcal{N}=2$ $SU(2)$ gauge 
theories and two-dimensional Virasoro CFT is generalized by Wyllard
to $SU(N)$ gauge theories \cite{Wyllard:2009hg}.
The instanton partition functions are then related with the conformal blocks 
for the non-linear conformal algebra $W_N$.
$W_2$-algebra is precisely the Virasoro symmetry of the original proposal.
In this section we study $SU(3)$ theories
in which case we know the explicit form of the commutation relations 
of the corresponding $W_3$-algebra.
Especially we focus on the $SU(3)$ theory with $N_f=1,2$ flavors and 
characterize the $N_f=1$ Whittaker state
as a simultaneous eigenstate of lowest annihilation operators.
As we will see in section 4 the case $N_f= N-1$ requires a redefinition of
the $W_N$ currents and we should take care of it in our definition of
the coherent state. 


\subsection{$\mathrm{W}_3$-algebra and the level one Shapovalov matrix}

$\mathrm{W}_N$-algebra is a kind of generalization of the Virasoro algebra.
This algebra is generated by the energy momentum tensor $T(z)$
and higher spin currents $W^{(s)}(z)$ $s=3,4,\cdots,N$.
This is the conformal symmetry of the Toda CFT of type $A_{N-1}$,
as the Virasoro symmetry controls the Liouville CFT.
In general, it is very hard to write down the explicit algebra of these currents with central extension.
Fortunately, the explicit form of the $\mathrm{W}_3$-algebra,
which is of our interest here, is known.
We first review the basic facts on $W_3$ algebra mainly following \cite{Mironov:2009by}
and fix our notations.
The $\mathrm{W}_3$-algebra consists of the energy-momentum tensor 
$T(z) = \displaystyle{\sum_{n \in {\mathbb{Z}}}z^{-n-2} L_n}$ 
and the spin-three current $W(z) = \displaystyle{\sum_{n \in {\mathbb{Z}}} z^{-n-3} W_n}$. 
The commutation relations among the modes $L_n, W_n$ are
\beqa
\left[ L_n, L_m \right] &=& (n-m) L_{n+m} + \frac{c}{12} (n^3 -n) \delta_{n+m, 0}, \\
\left[ L_n, W_m \right] &=& (2n-m) W_{n+m}, \\
\left[ W_n, W_m \right] &=& \frac{9}{2}\left[ (n-m) \left(
\frac{(n+m+2)(n+m+3)}{15} - \frac{(n+2)(m+2)}{6} \right) L_{n+m} \right. \CR
&&~~~~\left. + \frac{16}{22+5c} (n-m) \Lambda_{n+m} + \frac{c}{3\cdot 5!}n(n^2-1)(n^2-4)\delta_{n+m, 0},
\right],
\eeqa
where $\Lambda_n$ is a composite operator
\beq
\Lambda_n := \sum_{m \in \mathbb{Z}} : L_m L_{n-m} : + \frac{x_n}{5} L_n,
\eeq
with $x_{2\ell} = (1-\ell)(1+\ell), x_{2\ell+1} = (1-\ell)(12+\ell)$. 
We introduce $Q$ to parametrize the central charge $c = 2( 1- 12 Q^2)$.

Let us construct the highest weight representation of the conformal algebra.
The resulting representation space is called the Verma module of the algebra.
The highest weight state $\vert \Delta(\vec{\alpha}) \rangle$ of $W_3$-algebra is labeled by 
the Toda momenta $\vec{\alpha} = (\alpha, \beta)$ and satisfies
\beq
L_0 \vert \Delta(\vec{\alpha}) \rangle = \Delta(\vec{\alpha}) \vert \Delta(\vec{\alpha}) \rangle,
\qquad
W_0 \vert \Delta(\vec{\alpha}) \rangle = w(\vec{\alpha}) \vert \Delta(\vec{\alpha}) \rangle,
\eeq
where the eigenvalues are
\beqa
\Delta(\vec{\alpha}) := \alpha^2 + \beta^2 - Q^2,  \label{Wp1} \\
w(\vec{\alpha}) := \sqrt{\frac{4}{4-15Q^2}} \alpha (\alpha^2 - 3 \beta^2). \label{Wp2}
\eeqa
This highest weight state vanishes when we act the annihilation operators $L_{n>0}$ and $W_{n> 0}$.
The descendants in the Verma module,
which are generated by acting the creation operators,
are labeled by a pair of the Young diagrams
$\vec{Y} = \{ Y_L, Y_W \}$ as $L_{-{Y}_L} W_{-{Y}_W} \vert \Delta(\vec{\alpha}) \rangle$. 
Here we adopt the notation $L_{-{Y}}=L_{-Y_{d}}L_{-Y_{d-1}}\cdots L_{-Y_1}$.
With this basis of the Verma module, the Shapovalov matrix at level $N$ is
defined by the following Gram matrix
\beq
Q_{\Delta}^{(N)} (\vec{Y}; \vec{\widetilde{Y}}) := \langle \Delta(\vec{\alpha}) \vert 
W_{Y_W} L_{Y_L} L_{-\widetilde{Y}_L} W_{-\widetilde{Y}_W} \vert \Delta(\vec{\alpha}) \rangle.
\eeq
Since the matrix element $\langle \Delta(\vec{\alpha}) \vert 
W_{Y_W} L_{Y_L} L_{-\widetilde{Y}_L} W_{-\widetilde{Y}_W} \vert \Delta(\vec{\alpha}) \rangle$
is nonzero only for $|\vec{Y}| = |\vec{\widetilde{Y}}| =N$,
the full Shapovalov matrix $Q_{\Delta}$ is block-diagonal with the blocks $Q_{\Delta}^{(N)}$.
In the following let us denote the components of level one matrix and its inverse as
\beq
Q_{\Delta}^{(1)} =
\left(
\begin{array}{cc}
Q_{LL} & Q_{LW} \\
Q_{WL} & Q_{WW}
\end{array}
\right),
\qquad
(Q_{\Delta}^{(1)})^{-1} =
\left(
\begin{array}{cc}
R_{LL} & R_{LW} \\
R_{WL} & R_{WW}
\end{array}
\right),
\eeq
where the indices mean $L= \{ \unitbox\,, \phi \}$ and $W= \{ \phi, \unitbox\, \}$.
They are explicitly given by \cite{Mironov:2009by}
\beq
Q_{\Delta}^{(1)} =
\left(
\begin{array}{cc}
2 \Delta & 3 w \\
3 w & \frac{9 D \Delta}{2} 
\end{array}
\right),
\quad
(Q_{\Delta}^{(1)})^{-1} =
\frac{1}{9(D \Delta^2 - w^2)}
\left(
\begin{array}{cc}
 \frac{9 D \Delta}{2}  & -3 w \\
-3 w & 2 \Delta
\end{array}
\right),
\eeq
where 
\beq
D(\Delta) = \frac{4\Delta}{4-15 Q^2} + \frac{3 Q^2}{4 - 15 Q^2}.
\eeq
The level-$N$ Kac determinant  is given by the determinant of the Shapovalov matrix $\det Q_\Delta^{(N)}$.
At level one,
the Kac determinant can be factorized  as it should be;
\beq
D \Delta^2 - w^2 = \frac{4}{4-15 Q^2} \left( \beta^2 - \frac{Q^2}{4} \right)
((\beta - Q)^2 - 3 \alpha^2) (( \beta + Q)^2 - 3\alpha^2).
\eeq
If we use the following identification of parameters of CFT 
and SYM side\footnote{We have rescaled by $\sqrt{-\epsilon_1 \epsilon_2}$
to make the parameters on CFT side dimensionless.}  \cite{Mironov:2009by},
\beq
\alpha = \frac{\sqrt{3}}{2\sqrt{-\epsilon_1 \epsilon_2}} ( a_1 + a_2), \quad
\beta = \frac{1}{2\sqrt{-\epsilon_1 \epsilon_2}} ( - a_1 + a_2), \quad
Q =  \frac{\epsilon_1 + \epsilon_2}{\sqrt{-\epsilon_1 \epsilon_2}},
\eeq
we have
\beq
D \Delta^2 - w^2 =  \frac{-(a_{12} - \epsilon_{+}) (a_{12} + \epsilon_{+})
(a_{23} - \epsilon_{+}) (a_{23} + \epsilon_{+})
(a_{31} - \epsilon_{+}) (a_{31} + \epsilon_{+})}{(\epsilon_1\epsilon_2)^2 
(4 \epsilon_1 \epsilon_2 +15 \epsilon_{+}^2)},
\eeq
where $a_{ij} := a_i - a_j$ and $\epsilon_{+} := \epsilon_1 + \epsilon_2$. 
Notice that this is proportional to the denominator of 
the one instanton part of $SU(3)$ Nekrasov partition function.
We use this property to determine Whittaker states up to one instanton.
We also have
\beq
\label{DandW}
\Delta = \frac{a_1^2 + a_1 a_2 + a_2^2 - \epsilon_{+}^2}{-\epsilon_1 \epsilon_2}, \qquad
w = \frac{3\sqrt{3}}{\sqrt{4 \epsilon_1 \epsilon_2 +15 \epsilon_{+}^2}} 
\frac{a_1 a_2 (a_1 + a_2)}{-i \epsilon_1 \epsilon_2}.
\eeq
As we will see these factors appear in the numerator of the partition function.


\subsection{Whittaker state of $W_3$ algebra}

Let us consider a Whittaker vector of $W_3$ algebra 
in the Verma module over the highest weight state 
$ \vert \Delta(\vec{\alpha}) \rangle$. 
Our final goal is to construct such states whose scalar product  gives the instanton partition function
\beq
\label{AGTrank2}
\sum_{k=0}^\infty Z^{(N_f)}_{SU(3),k}=
\langle G_{N_f-n} \vert G_n \rangle,\quad
\textrm{for}
\quad
0 \leq N_f- n, n \leq 2,
\eeq
where $k$ labels the instanton number.
See Appendix A for details of the instanton partition function.
Actually, the states in the right hand side of \eqref{AGTrank2} belong to the $SU(3)$ theory
with $n$ fundamentals and $N_f-n$ anti-fundamentals in our convention.
Since the instanton number on the gauge theory side corresponds to
the level in the Verma module \cite{Gaiotto:2009ma}, among the annihilation 
operators $L_{n>0}, W_{n>0}$ the relevant ones
for the one instanton partition function are $L_1$ 
and $W_1$. Hence we introduce a simultaneous eigenstate;
\beq
L_1 \vert q_L, q_W \rangle = q_L  \vert q_L, q_W \rangle, \qquad
W_1 \vert q_L, q_W \rangle = q_W  \vert q_L, q_W \rangle,
\eeq
and  expand it in the Verma module as
\beq
 \vert q_L, q_W \rangle =   \vert \Delta(\vec{\alpha}) \rangle
 + c_L L_{-1}  \vert \Delta(\vec{\alpha}) \rangle + c_W W_{-1}  \vert \Delta(\vec{\alpha}) \rangle
 + \cdots .
\eeq
Looking at $\langle \Delta(\vec\alpha) \vert L_{-1} \vert  q_L, q_W \rangle$ and 
$\langle \Delta(\vec\alpha) \vert W_{-1} \vert  q_L, q_W \rangle$,
we find the equations that determine the coefficients $c_{L,W}$:
\beqa
q_L &=&  Q_{LL} c_L + Q_{LW} c_W, \CR
q_W  &=& Q_{WL} c_L + Q_{WW} c_W.
\eeqa
Hence the norm up to level one is
\beq
\langle q_L', q_W' \vert q_L, q_W \rangle 
= 1 + q_L' R_{LL} q_L + q_L' R_{LW} q_W + q_W' R_{WL} q_L + q_W' R_{WW} q_W + \cdots.
\eeq
On the other hand the one instanton part of the $SU(3)$ Nekrasov function
takes the following form;
\beq
Z_{SU(3), k=1}^{(N_f)} (a_1, a_2, m_i ; \epsilon_1, \epsilon_2)
= \Lambda^{6 - N_f} 
\frac{z^{(N_f)} (a_1, a_2, m_i ; \epsilon_1, \epsilon_2)}{D (a_1, a_2 ; \epsilon_1, \epsilon_2)}.
\eeq
While the numerator $z^{(N_f)}$ depends on the number $N_f$ of matter hypermultiplets,
the denominator is independent of matter contents and given by
\beqa
D &=&  \epsilon_1 \epsilon_2 (a_{12} - \epsilon_{+}) (a_{12} + \epsilon_{+})
(a_{23} - \epsilon_{+}) (a_{23} + \epsilon_{+})
(a_{31} - \epsilon_{+}) (a_{31} + \epsilon_{+}) \CR
&=&( - \epsilon_1 \epsilon_2)^3 (4 \epsilon_1 \epsilon_2  + 15 \epsilon_{+}^2 ) (D \Delta^2 - w^2).
\eeqa
If we have a matter in the fundamental representation with mass $m$
(see Appendix A for our convention of matter contribution to 
the instanton counting),
%
\beqa
z^{(1)} 
&=& -6(m - \frac{\epsilon_{+}}{2} ) ( a_1^2 + a_1 a_2 + a_2^2 - \epsilon_{+}^2) 
+ 9 a_1 a_2 (a_1 + a_2) \CR
&=& \epsilon_1 \epsilon_2 \left[ 27  (m - \frac{\epsilon_{+}}{2} ) R_{WW}
 + i \sqrt{27(4 \epsilon_1 \epsilon_2 + 15 \epsilon_{+}^2)} R_{WL} \right] (D \Delta^2 - w^2) .
\eeqa
Here we use (\ref{DandW}) to get the second equality.
Hence
\beq
\label{oneflavor}
Z_{SU(3), k=1}^{(1)} 
= \frac{\Lambda^5}{(\epsilon_1 \epsilon_2)^2(4 \epsilon_1 \epsilon_2 + 15 \epsilon_{+}^2) }
\left[  27  ( \frac{\epsilon_{+}}{2} - m ) R_{WW}
 - i \sqrt{27(4 \epsilon_1 \epsilon_2 + 15 \epsilon_{+}^2)} R_{WL} \right],
\eeq
is the one instanton partition function in the CFT language.
Comparing it with
\beq
\langle 0, q_0 \vert q_L, q_W \rangle = 1 \pm \sqrt{\frac{27}{4 \epsilon_1 \epsilon_2 + 15 \epsilon_{+}^2}} 
\frac{\Lambda^3}{\epsilon_1 \epsilon_2} (R_{WL} q_L + R_{WW} q_W) + \cdots,
\eeq
we find
\beq
q_L = \mp i \frac{ \Lambda^2}{\epsilon_1 \epsilon_2}, \qquad
q_W = \pm \sqrt{\frac{27}{4 \epsilon_1 \epsilon_2 + 15 \epsilon_{+}^2}}  
\frac{(\frac{\epsilon_{+}}{2}  -m ) \Lambda^2}{\epsilon_1 \epsilon_2},
\eeq
where we take $q_0 := \pm \sqrt{ \frac{27}{4 \epsilon_1 \epsilon_2 + 15 \epsilon_{+}^2} } 
\frac{\Lambda^3}{ \epsilon_1 \epsilon_2} $ for the Whittaker state of the pure $SU(3)$ theory \cite{Taki:2009zd}.
The $N_f=1$ Whittaker state is then $\vert G_1, m \rangle=\vert q_L, q_W\rangle$.
For anti-fundamental matter by replacing $m \to \epsilon_{+} -m$ we have
\beq
\widetilde{q_L }= q_L, \qquad
\widetilde{q_W} = \pm \sqrt{\frac{27}{4 \epsilon_1 \epsilon_2 + 15 \epsilon_{+}^2}}  
\frac{(m - \frac{\epsilon_{+}}{2}) \Lambda^2}{\epsilon_1 \epsilon_2}.
\eeq
From the viewpoint of the decoupling construction to be discussed in the next section,
it is natural to associate the Whittaker ket-vector 
$\langle G_1, m \vert=\langle {q_L }, \widetilde{q_W} \vert$ with anti-fundamental matter.


\subsection{ $N_f = 2$ theory} 

When $N_f = 2$ we have to consider two cases; in the symmetric realization 
one is fundamental and the other is anti-fundamental. In the asymmetric realization 
both are in the fundamental representation. 
These two cases correspond to $n=1$ and $n=0,2$ of (\ref{AGTrank2}) respectively.
The numerator of the one instanton partition function is
\beqa
z^{(2)}_{A} &=&
6(a_1^2 + a_1 a_2 + a_2^2 - \epsilon_{+}^2) (m_1 - \frac{\epsilon_{+}}{2}) (m_2 - \frac{\epsilon_{+}}{2}) 
-9 a_1 a_2 (a_1 + a_2) (m_1 + m_2 - \epsilon_{+} ) \CR
&& ~~ + 2(a_1^2 + a_1 a_2 + a_2^2)^2
- \frac{5}{2} (a_1^2 + a_1 a_2 + a_2^2) \epsilon_{+}^2 + \frac{1}{2} \epsilon_{+}^4,
\eeqa
for the asymmetric choice. By substituting $m_2 \to \epsilon_{+} - m_2$ we obtain
\beqa
z^{(2)}_{S} &=&
- 6(a_1^2 + a_1 a_2 + a_2^2 - \epsilon_{+}^2) (m_1 - \frac{\epsilon_{+}}{2}) (m_2 - \frac{\epsilon_{+}}{2}) 
-9 a_1 a_2 (a_1 + a_2) (m_1 - m_2) \CR
&& ~~ + 2(a_1^2 + a_1 a_2 + a_2^2)^2
- \frac{5}{2} (a_1^2 + a_1 a_2 + a_2^2) \epsilon_{+}^2 + \frac{1}{2} \epsilon_{+}^4,
\eeqa
for the symmetric choice. 
Using (\ref{DandW}), we can rewrite the first line of the equation into a CFT quantity.
The second line also has the following simple expression;
\beqa
(D \Delta^2 - w^2) R_{LL} 
&=& \frac{1}{2(4 -15 Q^2)} ( 4 (\alpha^2 + \beta^2)^2 - 5 Q^2 (\alpha^2 + \beta^2) + Q^4) \\
&=& \frac{2 (a_1^4 + a_2^4 + 2 a_1^3 a_2 + 2 a_1 a_2^3 + 3 a_1^2 a_2^2)
- \frac{5 \epsilon_{+}^2}{2} (a_1^2 + a_2^2 + a_1 a_2) + \frac{1}{2} \epsilon_{+}^4 }
{\epsilon_1 \epsilon_2 (4\epsilon_1 \epsilon_2 +15 \epsilon_{+}^2)} \nonumber.
\eeqa
Hence, for symmetric realization, the one instanton partition function is
\beqa
Z_{SU(3), k=1}^{(2)}  
&=& \left[ \frac{-27}{ 4 \epsilon_1 \epsilon_2 + 15 \epsilon_{+}^2 }
 (m_1 - \frac{\epsilon_{+}}{2}) (m_2 - \frac{\epsilon_{+}}{2}) R_{WW}   \right. \CR
&& ~~ \left. + i \sqrt{\frac{27}{4 \epsilon_1 \epsilon_2 + 15 \epsilon_{+}^2}} 
( m_1 - m_2)R_{LW}  - R_{LL} \right]  \frac{\Lambda^4}{(\epsilon_1 \epsilon_2)^2} \\
&=& \widetilde{q_W}(m_2) R_{WW} q_W(m_1) + \widetilde{q_W}(m_2) W_{WL} q_L + q_L W_{LW} q_W(m_1)
+ q_L R_{LL} q_L, \nonumber
\eeqa
and we have
\beq
1 + Z_{SU(3), k=1}^{(2)}  +\mathcal{O}(\Lambda^8)
= \langle q_L, \widetilde{q_W}(m_2) \vert q_L, q_W(m_1) \rangle
\eeq
at one instanton level. 
For the anti-symmetric realization we should have
\beqa
&&Z_{SU(3), k=1}^{(2)} 
= \langle 0, q_0 \vert q_L^{(2)}, q_W^{(2)} \rangle\vert_{\textrm{level}\, 1}
= \pm \frac{\epsilon_1 \epsilon_2 \Lambda^3}{D(a_1, a_2 ; \epsilon_1, \epsilon_2)}\CR
&&\quad\times \left(9 a_1 a_2 (a_1+a_2) q_L^{(2)} 
 + 6(a_1^2 + a_1 a_2 + a_2^2 - \epsilon_{+}^2)
  \sqrt{\frac{4 \epsilon_1 \epsilon_2 + 15 \epsilon_{+}^2}{27}} q_W^{(2)} \right),
\eeqa
which implies
\beqa
q_L^{(2)}(m_1, m_2)  &=& \mp (m_1 + m_2 - \epsilon_{+}) \frac{\Lambda}{\epsilon_1 \epsilon_2}, \label{qL}\\
q_W^{(2)} (m_1, m_2) &=& \pm \left[ (m_1 - \frac{\epsilon_{+}}{2})(m_2 - \frac{\epsilon_{+}}{2}) + \frac{1}{3} (a_1^2 + a_1 a_2 + a_2^2)
- \frac{1}{12} \epsilon_{+}^2 \right]  
\sqrt{\frac{27}{4 \epsilon_1 \epsilon_2 + 15 \epsilon_{+}^2}} \frac{\Lambda}{\epsilon_1 \epsilon_2}. \label{qW}\CR
&&
\eeqa
Note that $q_W^{(2)}$ depends on the Coulomb moduli $a_{1,2}$.
The $N_f=2$ Whittaker state up to level one is therefore $\vert G_2\rangle_{0+1}=\vert q_L^{(2)}, q_W^{(2)} \rangle$.
By substituting $m_{i} \to \epsilon_{+} - m_i$ we obtain the eigenvalues for the anti-fundamental matters;
\beq
\widetilde{q_L}^{(2)}(m_1, m_2)  = - q_L^{(2)}(m_1, m_2), \qquad 
\widetilde{q_W}^{(2)} (m_1, m_2) = q_W^{(2)} (m_1, m_2).
\eeq
${}_{0+1}\langle G_2\vert=\langle \widetilde{q_L}^{(2)}, \widetilde{q_W}^{(2)}\vert$ is the Whittaker state
for two anti-fundamental matters.
In the next section we will see that $\vert q_L^{(2)}, q_W^{(2)} \rangle$
cannot agree with the true Whittaker-like state $\vert G_2, m_1, m_2 \rangle$. The discrepancy 
appears beyond level one.
The genuine state $\vert G_2, m_1, m_2 \rangle$ is actually not an eigenstate of $W_1$,
but we can characterize it as a typical example of generalized Whittaker state.

We can check that the one instanton partition function with $N_f = 3,4$ can be reproduced from
the Whittaker states we have obtained. Namely 
\beqa
&& 1 + Z_{SU(3), k=1}^{(3)} +\mathcal{O}(\Lambda^6)=
 \langle q_L, \widetilde{q_W}(m_3) \vert q_L^{(2)}(m_1, m_2) , q_W^{(2)} (m_1, m_2) \rangle,  \\
&& 1 + Z_{SU(3), k=1}^{(4)} - \frac{1}{3} \frac{\Lambda^2}{\ep\es} +\mathcal{O}(\Lambda^4) \CR
&& ~~~= \langle \widetilde{q_L}^{(2)}(m_3, m_4), \widetilde{q_W}^{(2)} (m_3, m_4) 
\vert q_L^{(2)}(m_1, m_2) , q_W^{(2)} (m_1, m_2) \rangle .
\eeqa
When $N_f=4$, we observe a shift $- \frac{1}{3}\frac{\Lambda^2}{\ep\es}$, 
which arises from a remnant of the $U(1)$ contribution
$\exp(- \frac{1}{3} \frac{\Lambda^2}{\ep\es} )$. 



\subsection{Comment on $N_f=0$ theory}

The decoupling limit $m\to\infty$ and $m\Lambda^2\to\Lambda^3$ of
\eqref{oneflavor} leads to the Whittaker state $\vert G_0\rangle:= \vert 0, q_0 \rangle$
for the pure $SU(3)$ Yang-Mills theory.
This is an eigenstate of $W_1$
\beq
W_1 \vert G_0 \rangle =
\pm \sqrt{ \frac{27}{4 \epsilon_1 \epsilon_2 + 15 \epsilon_{+}^2} } 
\frac{\Lambda^3}{ \epsilon_1 \epsilon_2}
  \vert G_0\rangle,
\eeq
and annihilated by $L_{n\geq1}$ and $W_{n\geq2}$.
The solution to the condition is given by \cite{Taki:2009zd} as
\begin{align}
\label{pYMcoherent}
|G_0\rangle =
\sum_{\vec{Y}}
\left(\pm \sqrt{ \frac{27}{4 \epsilon_1 \epsilon_2 + 15 \epsilon_{+}^2} } 
\frac{\Lambda^3}{ \epsilon_1 \epsilon_2}\right)^{n}
Q^{-1}_{\Delta({\vec{\alpha}})}(
\varPhi ,[1^{n}];\vec{Y})
|\,\vec{Y}\, \rangle.
\end{align}
In the next section we will generalize this expression for the Whittaker states with $N_f=1,2$ flavors.
The expression (\ref{pYMcoherent}) also appeared in \cite{Keller:2011ek},
however the pre-factor $\sqrt{{27}/{4 + 15 (\epsilon_{+}^2/\epsilon_1 \epsilon_2)}}$ was
missing in the eigenvalue.
This is because they use another normalization of the generators $W_n$.
Their normalization is natural in the free-field construction of $W$-algebra
and this convention would be useful to extend our argument to generic gauge groups.



\section{Decoupling limit and generalized Whittaker state}

In the last section we assumed the existence of a Whittaker state of $W_3$ algebra 
and determined simultaneous eigenvalues of $L_1$ and $W_1$ by comparing the scalar product with
the one instanton partition function of $SU(3)$ theory with $N_f \leq 4$. 
In this section we will investigate the agreement at higer levels or instanton numbers 
by identifying the Whittaker state in the decouling limit from the superconformal theory 
with $N_f=6$, where the AGT-W relation tells that the Nekrasov partition function agrees 
with the conformal block of the Toda theory of $A_2$ type\footnote{{
Note that the agreement is still a conjecture and we asuume it in this section.}}. 
From the commutation relations of $W_3$ algebra
one may argue that additional non-vanishing eigenvalue is allowed only for $L_2$ 
even at higher levels, since $L_{n \geq 3}$ and $W_{m \geq 2}$ are written 
as (multiple) commutators of $L_1,L_2$ and $W_1$. We will see this is the case
for $N_f=1$ (actually the eigenvalue of $L_2$ vanishes). However, it turns out that
the state for $N_f=2$ should have non-vanishing eigenvalues of $W_2$ and $W_3$.
In this sense the $N_f=2$ Gaiotto-like state in $SU(3)$ theory cannot be a genuine Whittaker
state and we will call it generalized Whittaker state.
An ingenious mechanism to give non-zero eigenvalues of 
$W_2$ and $W_3$ will be made clear by studying $W_1$ action on the state.
In the next section we discuss the origin of non-vanishing eigenvalues 
of $W_2$ and $W_3$ from the viewpoint of Seiberg-Witten curve.


\subsection{Decoupling limit of ${W}_3$ conformal block}

We consider the $SU(3)$ theory with $N_f=6$ flavors. 
Following \cite{Mironov:2009dr}, we take six hypermultiplets
in anti-fundamental representation\footnote{This is just to make our argument parallel to that of \cite{Mironov:2009dr}.
We can easily get the answer for fundamental representation by replacing mass $\mu$ with $\epsilon_+-\mu$.}.
In this case the AGT-W relation is a correspondence between 
this superconformal $SU(3)$ theory and the $A_2$-type Toda CFT on the Riemann sphere with four punctures.
Two $A_2$-Toda momenta $\vec{\alpha}_{1,2}=(\alpha_{1,2},\beta_{1,2})$
and three mass parameters $\mu_{1,2,3}$ for the anti-fundamentals are related through
the dictionary of Wyllard that generalizes the AGT relation \cite{Alday:2009aq};
\begin{align}
\mu_1&=
-\frac{1}{\sqrt{3}}\alpha_1+\frac{Q}{2}+\frac{2}{\sqrt{3}}\alpha_2,
\nonumber\\
\mu_2&=
-\frac{1}{\sqrt{3}}\alpha_1+\frac{Q}{2}-\frac{1}{\sqrt{3}}\alpha_2-\beta_2,
\nonumber\\
\label{FMvsTM}
\mu_3&=
-\frac{1}{\sqrt{3}}\alpha_1+\frac{Q}{2}-\frac{1}{\sqrt{3}}\alpha_2+\beta_2.
\end{align}
Notice that in this section all the gauge theory parameters are dimensionless
by scaling out their overall mass scale with $\sqrt{-\epsilon_1\epsilon_2}$
as $\epsilon_+/\sqrt{-\epsilon_1\epsilon_2}= Q$.
In \eqref{FMvsTM} we choose $\vec{\alpha}_1$ to be the simple puncture of \cite{Gaiotto:2009we},
and the corresponding primary state will be of semi-null type \cite{Wyllard:2009hg} 
with momentum $\beta_1=-Q/2$.
See \cite{Wyllard:2009hg, Mironov:2009by} for details.
The anti-fundamental masses $\mu_{4,5,6}$ are given by the 
remaining Toda momenta $\vec{\alpha}_{3,4}$ through a similar relation:
\begin{align}
\mu_4&=
\frac{1}{\sqrt{3}}\alpha_3+\frac{Q}{2}+\frac{2}{\sqrt{3}}\alpha_4,
\nonumber\\
\mu_5&=
\frac{1}{\sqrt{3}}\alpha_3+\frac{Q}{2}-\frac{1}{\sqrt{3}}\alpha_4-\beta_4,
\nonumber\\
\label{AFMvsTM}
\mu_6&=
\frac{1}{\sqrt{3}}\alpha_3+\frac{Q}{2}-\frac{1}{\sqrt{3}}\alpha_4+\beta_4.
\end{align}
Now $\vec{\alpha}_3$ corresponds to the simple puncture.
The proposal of AGT-W is that the instanton partition function
is precisely equal to the $W_3$ conformal block of the sphere with four punctures $\vec{\alpha}_{1,\cdots,4}$:
\begin{align}
Z_{SU(3)}^{N_f=6}(a_1,a_2,\mu_{i},q;\epsilon_1,\epsilon_2)=Z_{U(1)}
\sum q^{|\vec{Y}|}
\langle  V_3V_4 V_{\vec{Y},\vec{\alpha}}\rangle
\,Q^{-1}_{\Delta({\vec{\alpha}})}(\vec{Y};\vec{Y}')\,
\langle V_{\vec{Y}'\vec{\alpha}} \vert V_1V_2 \rangle,
\end{align}
where $Q^{-1}_{\Delta({\vec{\alpha}})}$ denotes 
the inverse of the Shapovalov matrix. Here the building blocks
$\langle  VV V_{\vec{Y},\vec{\alpha}}\rangle$ and $\langle  V_{\vec{Y},\vec{\alpha}}|VV \rangle$ 
are the spherical three point conformal blocks
which include a descendant field insertion. 
See \cite{Mironov:2009dr} for more details.

Let us consider the decoupling limit of a single anti-fundamental hypermultiplet $\mu_1$
and three anti-fundamentals $\mu_{4,5,6}$.
This is done by taking infinitely-massive limit of these matter fields.
To get a non-empty theory after this limit,
we also have to scale the gauge coupling constant of the theory correctly.
Thus the decoupling limit to $N_f=2$ theory is
\begin{align}
\mu_{1,4,5,6}\rightarrow \infty,\quad
\mu_{2,3}\,:\,\textrm{fixed},\quad
\textrm{with}\quad q\,\mu_1\,\mu_4\,\mu_5\,\mu_6\,\to (\Lambda_{N_f=2})^4,
\end{align}
where $q=e^{2\pi i\tau}$ is the coupling constant of $N_f=6$ theory.
$\Lambda_{N_f=2}$ is the dynamical scale of the gauge theory with two flavors. 
We can translate this procedure into a scaling limit of the conformal block
through the dictionary (\ref{FMvsTM}) and (\ref{AFMvsTM}).
Since the full-decoupling limit $\mu_{4,5,6}\rightarrow \infty$ of the anti-fundamentals
was discussed in detail in \cite{Taki:2009zd},
we focus on the decoupling limit of the hypermultiplet $\mu_1\rightarrow \infty$.
In the language of Toda CFT, this limit means
\begin{align}
&C:=\alpha_1+\alpha_2=\frac{\sqrt{3}}{2}\left(Q-(\mu_2+\mu_3) \right)
\,:\,\textrm{fixed},\\
&2\beta_2=-\mu_2+\mu_3
\,:\,\textrm{fixed},\\
&A:=\alpha_1\rightarrow\infty.
\end{align}
From \eqref{Wp1} and \eqref{Wp2} the conformal dimensions for the external vertex operators are given by
the Toda momenta as
\begin{align}
&\Delta_1=A^2-\frac{3}{4}Q^2,\qquad\qquad\, \Delta_2=(C-A)^2+{\beta_2}^2-Q^2,
\\
&w_1=\sqrt{\kappa}\,A\left(A^2-\frac{3}{4}Q^2\right)
,\quad w_2=\sqrt{\kappa}\,(C-A)\left((C-A)^2-3{\beta_2}^2\right)
,
\end{align}
where $\sqrt{\kappa}:=\sqrt{\frac{4}{4-15Q^2}}$.
From these expressions we can determine asymptotic values of the three point functions $\langle VVV\rangle$
in the decoupling limit $A\to\infty$.

The key ingredient in our computation is 
the recursion relations for the three point conformal blocks which were developed
by Russian group \cite{Mironov:2009dr}:
\begin{align}
\label{recursionL1}
&\langle L_{-n} V_{
\vec{Y},\vec{\alpha}
}| V_1(1)V_2(0)\rangle
=(\Delta_{\vec{Y},\vec{\alpha} } +n\Delta_{1}-\Delta_{2})
\langle V_{\vec{Y},\vec{\alpha}}| V_1(1)V_2(0) \rangle,\\
\label{recursionW1}
&\rule{0pt}{5ex}
\langle W_{-n} V_{\vec{Y},\vec{\alpha}}| V_1(1)V_2(0)\rangle
=\langle W_0V_{\vec{Y},\vec{\alpha}}| V_1(1)V_2(0) \rangle
+\left(\frac{n(n+3)w_{1}}{2}-w_{2}\right)
\langle V_{\vec{Y},\vec{\alpha}}| V_1(1)V_2(0) \rangle\nonumber\\
&\hspace{9cm} +n\langle V_{\vec{Y},\vec{\alpha}}| (W_{-1}V_1)(1)V_2(0) \rangle,
\end{align}
where $\vec{Y},\vec{\alpha}$ is the label for the descendants.
Since we choose the external state $V_1$ as a semi-null state, 
the action of $W_{-1}$ on the primary is given by
$W_{-1}V_1=\frac{3w_1}{2\Delta_1}L_{-1}V_1$
and we obtain
\begin{align}
\label{3rdterm}
\langle V_{\vec{Y},\vec{\alpha}}| (W_{-1}V_1)(1)V_2(0) \rangle
&=\frac{3w_1}{2\Delta_1}
\langle V_{\vec{Y},\vec{\alpha}}| (L_{-1}V_1)(1)V_2(0) \rangle\nonumber\\
&\rule{0pt}{4ex}
=\frac{3w_1}{2\Delta_1}
(\Delta_{\vec{Y},\vec{\alpha}}-\Delta_1-\Delta_2)
\langle V_{\vec{Y},\vec{\alpha}}| V_1(1)V_2(0) \rangle
.
\end{align}
Our current interest is in the behavior of these three point functions in the limit $\mu_1\to\infty$.
On the CFT side, this is the heavy momenta limit $\alpha_{1,2}\to\pm\infty$ with fixed $C$.
Since the external momenta are very large $\Delta_{1,2},w_{1,2}>>\Delta_{\vec{Y},\vec{\alpha}}$,
the asymptotic behavior of these three point function is then given by
\begin{align}
&\langle L_{-n} V_{\vec{Y},\vec{\alpha}}| V_1(1)V_2(0)\rangle
\sim (n\Delta_{1}-\Delta_{2})
\langle V_{\vec{Y},\vec{\alpha}}| V_1(1)V_2(0) \rangle, \label{asymL} \\
&\rule{0pt}{4ex}
\langle W_{-n} V_{\vec{Y},\vec{\alpha}}| V_1(1)V_2(0)\rangle \nonumber\\
&\rule{0pt}{4ex}
\qquad\quad\sim
\left(\frac{n(n+3)w_{1}}{2}-w_{2}
-n
\frac{3w_1}{2\Delta_1}
(-\Delta_{\vec{Y},\vec{\alpha}}+\Delta_1+\Delta_2)\right)
\langle V_{\vec{Y},\vec{\alpha}}| V_1(1)V_2(0) \rangle. \label{asymW}
\end{align}
In the following, we show that the dominant contributions in the decoupling limit come from
$L_{-1,-2}$ and $W_{-3,-2,-1}$.
The asymptotic values of the following factors
in \eqref{asymL} and \eqref{asymW} are crucial in our analysis:
\begin{align}
f_n =(n\Delta_{1}-\Delta_{2}),\quad
g_n
=\left(\frac{n(n+3)w_{1}}{2}-w_{2}
-n
\frac{3w_1}{2\Delta_1}
(-\Delta_{\vec{Y},\vec{\alpha}}+\Delta_1+\Delta_2)\right).
\end{align}
It is easy to see the following behavior of $f_n$ in the limit;
\begin{align}
f_1\sim 2CA,\quad
f_2\sim A^2,\quad
f_{n(\geq3)}< \mathcal{O}(A^3).
\end{align}
Among the Virasoro descendants $L_{-Y_L}V_{{Y}_W,\vec{\alpha}}$ with a fixed level $\ell=|{Y}_L|$, 
the special one $L_{-2}^rL_{-1}^sV_{Y_{W},\vec{\alpha}}$ therefore gives
the dominant contribution $A^{2r+s}=A^\ell$ to the conformal block in the decoupling limit.
Similarly the behavior of the $W$-descendants is controlled by
\begin{align}
& g_1\sim \sqrt{\kappa}\left( \frac{3}{2}C^2-\frac{9}{2}{\beta_2}^2 +\frac{9}{8}Q^2+\frac{3}{2}
(\Delta_{\vec{\alpha}}+\vert\vec{Y}\vert)\right)A,\quad
g_2\sim 3C\sqrt{\kappa} A^2,\nonumber\\
& g_{3}\sim \sqrt{\kappa} A^3,\quad
g_{n(\geq4)}< \mathcal{O}(A^4),
\end{align}
and then the descendants $W_{-3}^pW_{-2}^qW_{-1}^{t}V_{\vec{\alpha}}$
give the dominant contribution to the three point functions
for a fixed $|Y_W|$.
By combining there results, we find the following descendants dominate in the level $n$:
\begin{align}
\langle\,L_{-2}^rL_{-1}^sW_{-3}^pW_{-2}^qW_{-1}^{n-3p-2(q+r)-s}V_{\vec{\alpha}} \,|\,V_1\,V_2\,\rangle
\sim
\textrm{const.}\,A^n.
\end{align}
These are the three point functions for the descendants labeled by 
$Y_L=[2^r\cdot 1^s],\,Y_W=[3^p\cdot 2^q\cdot 1^{n-3p-2(q+r)-s}]$.
Therefore only the contributions from these indices survive in the scaling limit of the conformal block.
The decoupling limit of the anti-fundamental hypermultiplets
$\mu_{4,5,6}\to\infty$ implies the following asymptotic value of the remaining three point function \cite{Taki:2009zd}:
\begin{align}
\langle  V_3V_4 V_{\vec{Y}',\vec{\alpha}}\rangle
\sim (\mu_4\mu_5\mu_6)^n
\frac{
(\sqrt{3})^{3n}}
{(\sqrt{4-15Q^2})^n}
\delta_{\vec{Y}';\varPhi,[1^n]}.
\end{align}
This dominant term comes from the special descendant $Y_L'=\varPhi,\,Y_W'=[1^n]$.
In this way we find that the irregular conformal block for $N_f=2$ takes the form\footnote{
The $U(1)$ factor $Z_{U(1)}$ gives a trivial contribution in the decoupling limit.}
\begin{align}
\label{iregCB}
\mathcal{B}^{(N_f=2)} :&= \lim
\sum q^{|\vec{Y}|}
\langle  V_3V_4 V_{\vec{Y}',\vec{\alpha}}\rangle
\,Q^{-1}_{\Delta({\vec{\alpha}})}(\vec{Y}';\vec{Y})\,
\langle V_{\vec{Y},\vec{\alpha}}\vert V_1V_2 \rangle  \nonumber\\
&\rule{0pt}{5ex}
=
\sum_{\substack{n,p,q,r,s \geq 0\\ n\geq 3p+2q+2r+s}}
 (-\Lambda_{N_f=2}^4)^{n}\,
\frac{2^{-n+4p+2q+2r+s}
(\sqrt{3})^{6n-12p-5q-8r-3s}}
{(\sqrt{4-15Q^2})^{2n-2p-q-2r-s}}
\nonumber\\
&
\qquad\times
q_L(\mu_2, \mu_3)^{s+q}\,
\prod_{\ell=0}^{n-3p-2q-2r-s-1}\left(q_W(\mu_2, \mu_3)+\frac{2\ell}{3}\right)
\nonumber\\
&
\qquad\times
Q^{-1}_{\Delta({\vec{\alpha}})}\,(
[2^r\cdot 1^s],[3^p\cdot 2^q\cdot 1^{n-3p-2(q+r)-s}]
;\,\varPhi, [1^n]),
\end{align}
where
\begin{align}
q_L(\mu_2,\mu_3) &:= Q - \mu_2 - \mu_3,
\\
q_W (\mu_2, \mu_3)  &:= 2 \mu_2 \mu_3 - Q (\mu_2 + \mu_3) + Q^2 + \frac{2}{3} (a_1^2 + a_1 a_2 + a_2^2 - Q^2).
\end{align}
Up to overall normalization these are nothing but $q_{L,W}^{(2)}$ in the previous section.
Notice that we use $\mu_1\sim -\sqrt{3}A$ to take the decoupling limit.
The above formula is an explicit expression of the irregular conformal block
for $N_f =2$ theory.
The level-one part of the irregular conformal block is
\begin{align}
\mathcal{B}^{(N_f=2)}_{\,\,\,1} = -
\frac{
27 \Lambda^4}
{2({4-15Q^2})} q_W (\mu_2, & \mu_3)
\,Q^{-1}_{\Delta({\vec{\alpha}})}\,(
\varPhi,[1]
;\,\varPhi, [1])
\nonumber\\
&\rule{0pt}{5ex}-
\frac{
{3}\sqrt{3} \Lambda^4}
{\sqrt{4-15Q^2}} q_L(\mu_2,\mu_3)
\,Q^{-1}_{\Delta({\vec{\alpha}})}\,(
[1],\varPhi
;\,\varPhi, [1]).
\end{align}
In Appendix B,
we show that this function actually reproduces the one-instanton partition function of
$SU(3)$ gauge theory with $N_f=2$ flavors.


\subsection{$N_f \leq 2$ Whittaker states}
Recall that the irregular conformal block for the $SU(3)$  theory with $N_f=2$ anti-fundamental flavors
is the scalar product of the Whittaker states for $N_f=2$ and $N_f=0$;
\beq
\mathcal{B}^{(N_f=2)}=
\langle G_2, m_1,m_2
\vert G_0
\rangle.
\eeq
The expression for the irregular conformal block (\ref{iregCB}) therefore
tells the following Gaiotto-like states for ${W}_3$-algebra:
\begin{align}
\label{Gstate0}
&|\,G_0\, \rangle
=\sum_{\vec{Y}}
\,(i\Lambda^{3})^n
\left(
\frac{
3\sqrt{3}
}
{
\sqrt{4-15Q^2}
}
\right)^n
Q^{-1}_{\Delta({\vec{\alpha}})}(
\varPhi, [1^n];\vec{Y})
\cdot
|\,\vec{Y}\, \rangle,
\\
&|\,G_2,\, m_1,m_2\rangle
=\rule{0pt}{5ex}
\sum_{\substack{\vec{Y},p,q,r,s \geq 0\\ n\geq 3p+2q+2r+s}}
\,(i\Lambda)^{n}
\frac{2^{-n+4p+2q+2r+s}
(\sqrt{3})^{3n-12p-5q-8r-3s}}
{(\sqrt{4-15Q^2})^{n-2p-q-2r-s}}
\nonumber\\
&\rule{0pt}{5ex}\times \left(q_L(m_1, m_2)\right)^{s+q}
\left(\frac{2}{3}\right)^{n-3p-2q-2r-s}
\left(\frac{3\,q_W(m_1, m_2)}{2}\right)_{n-3p-2q-2r-s}\nonumber\\
&
\label{Gstate2}
\rule{0pt}{5ex}\times
Q^{-1}_{\Delta({\vec{\alpha}})}(
[2^r\cdot 1^s],[3^p\cdot 2^q\cdot 1^{n-3p-2(q+r)-s}];\vec{Y})
\cdot
|\,\vec{Y}\, \rangle,
\end{align}
where $(x)_n:=x(x+1)\cdots(x+n-1)$ is the Pochhammer symbol.
There is an ambiguity of overall sign $\pm$
in front of $\Lambda$ in the definition of Whittaker states
and in this section we choose $+$ sign for simplicity. 
From \eqref{Gstate2}  we obtain the following conditions for the $N_f=2$ Whittaker state $|G_2,m_1,m_2\rangle\,
(=:|G_2\rangle)$ for anti-fundamentals;
\begin{align}
L_1& |\,G_2\, \rangle = i\Lambda \left(Q-(m_1+m_2)\right) |\,G_2\, \rangle,\label{L1}\\
L_2& |\,G_2\, \rangle = \frac{(i\Lambda)^2}{3} |\,G_2\, \rangle,\\
\left( W_1 - wi \Lambda L_0 \right)&
 |\,G_2\,\rangle
 = \frac{3 wi\Lambda}{2} 
\left(2m_1m_2-Q(m_1+m_2)+Q^2 \right)  |\,G_2\,\rangle,\label{Wcond}\\
W_2 &|\,G_2\,\rangle =w (i\Lambda)^2
\left(Q-(m_1+m_2)\right) |\,G_2\, \rangle, \\
W_3& |\,G_2\,\rangle =  \frac{ 2w(i \Lambda)^3}{9}
 |\,G_2\, \rangle,
\end{align}
where 
\beq
w := \frac{\sqrt{3}}{\sqrt{4-15Q^2}} 
\eeq
and  $|G_2\rangle$ is annihilated by $L_{n \geq 3}$ and  $W_{m \geq 4}$.
Change the masses $m_{1,2}$ into $\epsilon_+-m_{1,2}$
to get the result for fundamental hypermultiplets.
See Appendix C for a derivation of these equations.
Actually as we will see below the last two conditions for $W_{2,3}$ follow from the first three conditions.
Note that,
up to one instanton, the conditions \eqref{L1} and \eqref{Wcond} agree with \eqref{qL} and \eqref{qW}, respectively.
To see this,
let us recall that the Whittaker state is the following superposition of the states in the Verma module
\begin{align}
|\,G_2\, \rangle =\sum_{n=0}^\infty \Lambda^{n}\,\vert n, \Delta \rangle,
\end{align}
where $\vert n, \Delta \rangle$ is explicitly given by \eqref{Gstate2}.
The action of $L_0$ involves therefore the following Euler derivative with respect to the dynamical scale
\begin{align}
L_0|\,G_2\, \rangle
=\left(\Delta+ \Lambda \frac{\partial}{\partial\Lambda}\right)|\,G_2\,\rangle,
\end{align}
together with the eigenvalue $\Delta$ of $L_0$ on the primary state $\vert 0, \Delta \rangle$. 
This $\Delta$-term is the origin of the Coulomb moduli dependence of the
``eigenvalue" of the level one Whittaker state $\vert q^{(2)}_L,q^{(2)}_W \rangle$:
\begin{align}
&W_1 |\,G_2\,\rangle = \frac{3\sqrt{3} i\Lambda}{2\sqrt{4-15Q^2}} \nonumber\\
&{\times\left(2m_1m_1-Q(m_1+m_2)+Q^2
+
\frac{2}{3}(a_1^2 + a_1 a_2 + a_2^2 - Q^2)+
\frac{2\Lambda}{3}\frac{\partial}{\partial\Lambda}
 \right)} |\,G_2\,\rangle.
\end{align}
The derivative term of the right hand side vanishes for one instanton expansion of the relation
$\frac{\partial}{\partial\Lambda} |G_2\rangle_0=0$.
Notice that the relation (\ref{Wcond}) has the derivative term, 
or the level counting operator $L_0$,
and this state is therefore not a usual Whittaker state in the strict sense.
This is the reason why we introduce the notion of generalized Whittaker state.
The appearance of  such an Euler differential is not a mere accident,
and we will encounter a quite similar state in section 5,
where $SU(2)$ theory with a surface operator is studied. 
Note that $L_0$ in (\ref{Wcond}) plays an important role for the existence of a simultaneous eigenstate
of $L_{1,2}$ and $W_{2,3}$.
At first sight, it seems that the eigenvalues of $W_{2,3}$ should be zero because of the
commutation relation $(2n-3)W_n=[L_{n-1},W_1]$.
However $ |G_2\rangle$ is not a genuine eigenstate of $W_1$,
and so the emerging $L_0$ term gives the following contribution for $n>1$
\begin{align}
W_n|\,G_2\,\rangle
&=\frac{\sqrt{3}i \Lambda}{(2n-3)\sqrt{4-15Q^2}}
[L_{n-1},L_0]|\,G_2\,\rangle\nonumber\\
&=\frac{(n-1)\sqrt{3}i \Lambda}{(2n-3)\sqrt{4-15Q^2}}L_{n-1}|\,G_2\,\rangle,
\end{align}
which leads to non-zero eigenvalue for $n=2,3$.
Therefore the generalized Whittaker state involving $L_0$ term
can be a simultaneous eigenstate of  $L_{1,2}$ and $W_{2,3}$.

Let us consider the further decoupling limit of the Whittaker state (\ref{Gstate2}) to $N_f =1$.
By applying the limit $m_2\to\infty,\,\,\Lambda_{N_f=2} m_2\to (\Lambda_{N_f=1})^2$,
the state (\ref{Gstate2}) reduces to
\begin{align}
\label{Gstate1}
|\,G_1,\, m\rangle =
\sum_{\vec{Y},0\leq s\leq n}
\, (i\Lambda_{N_f=1}^2)^{n}
&\frac{2^{-n+s}
(\sqrt{3})^{3n-3s}}
{(\sqrt{4-15Q^2})^{n-s}}
\left(-1 \right)^{s}
\left(2m-Q\right)^{n-s}\, \nonumber\\
&\qquad\quad\,\,
\times
Q^{-1}_{\Delta({\vec{\alpha}})}(
[1^s],[1^{n-s}] 
; \vec{Y})
\cdot
|\,\vec{Y}\, \rangle
\end{align}
Thus, in the case of $N_f=1$ the eigenvalues of $L_1$ and $W_1$, which have been 
fixed at level one in section 2, completely characterize the Whittaker state.
The coherent condition for $N_f=1$ no longer involves the Euler derivative
and $|G_1,\, m\rangle$ is therefore a conventional Whittaker state for $L_1$ and $W_1$.
In addition, the eigenvalues are independent of the Coulomb moduli parameters.
Finally the $N_f=0$ limit $m\to\infty,\,\,(\Lambda_{N_f=1})^2 m\to (\Lambda_{N_f=0})^3$
for the state (\ref{Gstate1}) reproduces the Whittaker state (\ref{Gstate0}) for the pure super Yang-Mills theory.
The definition (\ref{Gstate0}) is therefore consistent with that of the partner (\ref{Gstate2}) in
the decomposition of  $\mathcal{B}^{(N_f=2)}$.



\section{Whittaker states from the Seiberg-Witten curve}

 As we have seen in the last section $N_f=2$ Whittaker state of $SU(3)$ theory
has non-vanishing eigenvalues of $W_2$ and $W_3$. 
In this section by looking at the Seiberg-Wiiten curve we argue
that when $N_f=N-1$ we have to redefine the $W_N$ currents 
to remove the $U(1)$ current. The non-vanishing eigenvalues of
higher modes come from a contribution from the $U(1)$ part.

The Seiberg-Witten curve of $SU(N)$ theory with $N_f$ fundamental 
matter is \cite{Argyres:1995wt,Hanany:1995na}
\beq
P_N(\lam) = \Lambda^N z + \frac{\Lambda^{N- N_f} Q_{N_f}(\lam)}{z},
\eeq
where
\beq
P_N (\lam):= \lam^N - \sum_{k=0}^{N-2} u_{N-k} \lam^k, 
\qquad Q_{N_f}(z) := \prod_{\ell=1}^{N_f} (\lam+ m_\ell).
\eeq
If we substitute $\lam = xz$, we obtain
\beq
x^N = \sum_{k=0}^{N-2} u_{N-k} x^k z^{k-N} + \Lambda^N z^{1-N}
+ \Lambda^{N-N_f} z^{N_f - N -1} \prod_{\ell=1}^{N_f} (x+ \frac{m_\ell}{z}),
\label{SWcurve}
\eeq
which may be compared with the Gaiotto curve
\beq
x^N = \sum_{n=2}^{N} \phi_n(z) x^{N-n}.
\eeq

When $N_f =0$, our curve is 
\beq
x^N = u_{2} x^{N-2}z^{-2} + \cdots + u_{N-1} x z^{1-N}+ u_N z^{-N} + \Lambda^N z^{1-N} +  \Lambda^{N} z^{ - N -1} .
\eeq
Thus we find
\beq
\phi_2(z) = u_{2} z^{-2} , \quad \cdots \quad,  \phi_{N-1}(z) = u_{N-1} z^{1-N},
\eeq
and
\beq
\phi_N(z) = u_N z^{-N} + \Lambda^N z^{1-N} +  \Lambda^{N} z^{ - N -1} .
\eeq
Following Gaiotto \cite{Alday:2009aq,Gaiotto:2009ma},
we want to identify $\phi_n(z)$ with $\langle G_0 \vert W^{(n)} (z) \vert G_0 \rangle /
\langle G_0 \vert G_0 \rangle$ in the limit $\epsilon_{1,2}\to 0$,
where $\vert G_0 \rangle$ is a Whittaker state in the Verma module. The mode expansion of 
the spin $n$ current is $ W^{(n)} (z) = \displaystyle{ \sum_{m \in \mathbb{Z}} z^{-m-n} W_m^{(n)} }$.
Hence we should have
\beq
\label{WSlower}
\frac{\langle G_0 \vert W_0^{(\ell)}\vert G_0 \rangle }{\langle G_0 \vert G_0 \rangle}
=u_{ \ell}(\epsilon_1,\epsilon_2), \quad  W_n^{(\ell)} \vert G_0 \rangle = 0,
\qquad 2 \leq \ell \leq N-1, n >0,
\eeq
and\footnote{
Notice that the second condition of (\ref{WSpure}) implies
 $\langle G_0\vert W_1^{(\ell)} =\Lambda^N\langle G_0\vert $
 since the conjugation is now defined as $W_n^\dag=W_{-n}$.}
\beq
\label{WSpure}
\frac{\langle G_0 \vert W_0^{(N)}\vert G_0 \rangle }{\langle G_0 \vert G_0 \rangle}
=u_{ N}(\epsilon_1,\epsilon_2), \quad  W_1^{(N)} \vert G_0 \rangle = \Lambda^N \vert G_0 \rangle,
\eeq
where $u_{ \ell}(\epsilon_1,\epsilon_2)=u_{ \ell}+\mathcal{O}(\epsilon)$
 are the ``quantum corrected" Coulomb moduli in the presence of the $\Omega$ background. 
 Actually the conditions ${\langle G_0 \vert W_0^{(\ell)}\vert G_0 \rangle }/{\langle G_0 \vert G_0 \rangle}
=u_{ \ell}(\epsilon_1,\epsilon_2)$ in \eqref{WSlower} for the zero-modes 
are achieved by the fact that the Whittaker state $\vert G_0 \rangle$ 
belongs to the Verma module with the primary state 
$ W_0^{(\ell)}\vert\vec{\alpha} \rangle=w_0^{(\ell)}(\alpha)\vert\vec{\alpha} \rangle$ 
and the dictionary of the AGT-W relation. Note that 
the Whittaker state is a superposition of vectors in different levels of the Verma module and
it is not an eigenstate of the zero-modes.

When $N_f = N-1$ there appears an $x^{N-1}$ term in \eqref{SWcurve}. One can eliminate it by 
an appropriate shift of $x \to x +c$. 
Since the linear term describes the center-of-mass degrees of freedom in the brane construction,
this completing square is just  the decoupling of the overall $U(1)$ subgroup of $U(N)$ gauge group.
Hence it is natural to introduce an additional $U(1)$ current $W^{(1)}$
in our $W$ algebra. Since the coefficient of the $x^{N-1}$ term takes a universal form $\Lambda z^{-2}$,
the condition concerning the $U(1)$ current is actually independent of $N$;
\beq
W_1^{(1)} \vert G_{N-1} \rangle = \Lambda \vert G_{N-1} \rangle ,
 \qquad W_n^{(1)} \vert G_{N-1} \rangle = 0, \qquad  n > 1. \label{U1}
\eeq
Let us see how it works for $SU(2)$ case where the curve is\footnote{
The mass term in the eigenvalue is actually corrected by  $\epsilon_+$ dependent term
as we saw in the previous section.
We can fix it by comparing the state with one instanton partition function.
In this section, we neglect $\epsilon_+$ shift of mass parameters.
}
\beq
x^2 = \Lambda z^{-2} x +  m \Lambda z^{-3} + u z^{-2} + \Lambda^2 z^{-1}.
\eeq
The condition for the Gaiotto-Whittaker state $\vert G_1 \rangle$ for $N_f=1$ is
\begin{align}
\frac{\langle G_0 \vert L_0 \vert G_1 \rangle}{\langle G_0 \vert G_1 \rangle} = u, \quad
L_1 \vert G_1 \rangle = m \Lambda \vert G_1 \rangle.
\end{align}
and 
\beq
J_1 \vert G_{1} \rangle = \Lambda \vert G_{1} \rangle ,
\eeq
from \eqref{U1}. Now let us modify the original Virasoro by
\beq
\label{Lmodif}
\widetilde{L}(z) = L(z) + \alpha : J(z) J(z) :.
\eeq
Then the condition for the Gaiotto-Whittaker state in terms of the new Virasoro generator is
\beq
\widetilde{L}_1 \vert G_1 \rangle = m \Lambda\vert G_{1} \rangle ,
 \quad \widetilde{L}_2 \vert G_1 \rangle = \alpha \Lambda^2\vert G_{1} \rangle ,
\eeq
where we have {\it assumed} $J_0 \vert G_1 \rangle =0$. We recover the original form proposed by Gaiotto \cite{Gaiotto:2009ma}. 
Recall that $\hat{x}^2=\hat{x}\hat{J}(z)+\hat{L}(z)$ describes the ``quantum Seiberg-Witten curve" \cite{Alday:2009aq}
because the expectation value of this CFT operator gives the classical curve of the corresponding gauge theory.
The modification of the energy momentum tensor (\ref{Lmodif}) is thus equivalent to the completing square of the curve
$(\hat{x}+\hat{J}(z)/2)^2=(\hat{x}+\hat{J}(z)/2)\hat{J}(z)+\hat{L}(z)$.
Hence,  after decoupling the center-of-mass degrees of freedom,
the modification (\ref{Lmodif}) is induced and the Whittaker state becomes 
a simultaneous eigenstate of $L_{1,2}$. 

Let us do the same with $N=3$ and $N_f =2$. The curve is
\beq
x^3 = \Lambda z^{-2} x^2 + ( u z^{-2} + (m_1 + m_2) \Lambda z^{-3} ) x 
+ v z^{-3} + \Lambda^3 z^{-2} + m_1 m_2 \Lambda z^{-4}.
\eeq
The condition for the Whittaker state $\vert G_2 \rangle$ for $N_f=2$ is therefore
\beqa
&&\frac{\langle G_0 \vert L_0 \vert G_2 \rangle}{\langle G_0  \vert G_2 \rangle} = u(\epsilon), \quad
L_1 \vert G_2 \rangle = (m_1 + m_2)  \Lambda\vert G_2 \rangle , \\
&&\frac{\langle G_0 \vert W_0 \vert G_2 \rangle}{\langle G_0  \vert G_2 \rangle}  = v(\epsilon), \quad
W_1 \vert G_2 \rangle = m_1 m_2 \Lambda\vert G_2 \rangle , \quad
\eeqa
and again we put
\beq
J_1 \vert G_{2} \rangle = \Lambda \vert G_{2} \rangle.
\eeq
Now as before let us consider the following redefinition of the currents;
\beq
\widetilde{L}(z) = L(z) + \alpha : J(z) J(z) :, \quad
\widetilde{W}(z) = W(z) + \beta : L(z) J(z): + \gamma : J(z) J(z) J(z):.
\eeq
Assuming $J_0 \vert G_2 \rangle =0$, we find 
the condition for $\vert G_2 \rangle $ in terms of the new generators
\beqa
&&\widetilde{L}_1 \vert G_2 \rangle = (m_1 + m_2)  \Lambda \vert G_2 \rangle,
 \quad \widetilde{L}_2 \vert G_2 \rangle = \alpha \Lambda^2 \vert G_2 \rangle, \\
&&\widetilde{W}_1 \vert G_2 \rangle = (m_1 m_2 \Lambda +  \beta \Lambda \widetilde{L}_0) \vert G_2 \rangle, \\
&&\widetilde{W}_2 \vert G_2 \rangle = \beta  (m_1 + m_2)  \Lambda^2 \vert G_2 \rangle, 
\quad \widetilde{W}_3 \vert G_2 \rangle = \gamma \Lambda^3 \vert G_2 \rangle.
\eeqa
The eigenvalue of $\widetilde{W}_3$ must be $\gamma={2\alpha\beta}/{3}$
so that $\widetilde{L}_n$ and $\widetilde{W}_m$ form the closed $W_3$-algebra again.
In section 3, we observed that when $N_f=2$ the action of $W_1$ on the Whittaker state
involves the Virasoro zero mode which
leads to the moduli parameter $a_{1,2}$ dependence. We see
this fact comes from the remnant of $U(1)$ current. 
The parameters $\alpha, \beta, \gamma$ can be fixed by comparing it with the Nekrasov partition function,
or by estimating the decoupling limit from the superconformal theory as we have done in the last
section.
These parameters should also be fixed by completing cube of the original curve which eliminates the
quadratic term in $x$.
This is because the completed curve must be consistent with the original one including the $U(1)$
degrees of freedom.
We should emphasize that there are \lq\lq quantum corrections\rq\rq\ by the $\Omega$ background $\ep$ and $\es$. 
In principle, these $\epsilon$ dependences would be fixed by comparing the expectation values of currents
$W^{(n)}$ with those of gauge theory operators
$\textrm{tr}~\Phi^n$ in the presence of $\Omega$ background.



\section{$SU(2)$ theory with a surface operator}

In this section we consider the instanton partition function in the presence of the surface operator
which gives a codimension two defect. 
Such an operator is realized by imposing the boundary condition in performing the path integral as
\beq
A_\mu dx^\mu\sim\textrm{diag}(\alpha_1,\alpha_2,\cdots,\alpha_N)id\theta,
\eeq
near the insertion locus $(z= r e^{i\theta} = 0)$ of the defect \cite{Gukov:2006jk}.
We can then define the instanton partition function in the presence of surface operator
by integrating over the field configurations with this boundary condition.
In general there are several types of the surface operator according to the breaking pattern of
the gauge symmetry on the defect. 
The remaining gauge symmetry is a subgroup of $U(N)$
given as the commutant of $\textrm{diag}(\alpha_1,\alpha_2,\cdots,\alpha_N)$
which is classified by the partitions of $N$. 
In the following we only consider $U(2)$ or $SU(2)$ theory where we have a unique surface
operator corresponding to $U(2) \to U(1)^2$. When we introduce the surface operator, 
the moduli space is labeled by a new topological
number (monopole, or vortex number) in addition to the instanton number. Hence the instanton 
partition function is expanded in two parameters $x$ and $z$.

Accoriding to \cite{Alday:2009fs}  the surface operator in $\mathcal{N}=2$ gauge theory corresponds to
the degenerate primary field $\Phi_{1,2}(x)$ that has 
the conformal weight $(b+b^{-1})^2/4 - (b/2 + b^{-1})^2$
and is in the Virasoro Verma module with
the central charge $c = 1 + 6 ( b + b^{-1} )$.
The point is that $\Phi_{1,2}(x)$ satisfies 
the null state condition $(b^2 L_{-1}^2 + L_{-2}) \Phi_{1,2} =0$. 
Using the Gaiotto states $\vert G_0 \rangle$ and $\vert G_1, m \rangle$ that are
defined by the same conditions in the $SU(2)$ theory without the surface operator, we have
\beqa
\label{simpleS0}
Z_{SU(2)}^{(S), N_f=0} &=& \frac{\langle G_0, - \vert \Phi_{1,2}(x) \vert G_0, + \rangle} 
{\langle G_0 \vert G_0 \rangle},  \\
Z_{SU(2)}^{(S), N_f=1}  &=& \frac{\langle G_1, - , m \vert \Phi_{1,2}(x) \vert G_0, + \rangle} 
{\langle G_1, m \vert G_0 \rangle},  \\
Z_{SU(2)}^{(S), N_f=2}  &=& \frac{\langle G_1, -, m_1 \vert \Phi_{1,2}(x) \vert G_1, +, m_2 \rangle} 
{\langle G_1, m_1 \vert G_1, m_2 \rangle},
\eeqa
where $Z_{SU(2)}^{(S), N_f}$ denotes the instanton partition function with the surface operator.
Actually this description through the Virasoro degenerate field is believed to correspond to the
surface operator of simple type \cite{Alday:2009fs}.
Note that since there is a $\Phi_{1,2}(x)$ operator insertion in the numerator,
the states $\vert G_0, \pm \rangle$ and $\vert G_1, \pm,  m \rangle$ should be in the
Verma module over the primary field with the conformal weight $\Delta(a \pm \frac{1}{4b}) :=
(b+b^{-1}){^2}/4 - (a \pm 1/(4b))^2$. In this approach the mode expansion parameter $x$ of
the degenerate field $\Phi_{1,2}(x)$ gives the monopole expansion parameter. 
On the other hand it was shown in \cite{Alday:2010vg} that the same partition function can be
obtained from the conformal block of the affine $\mathfrak{sl}(2)$ algebra.
This affine CFT describes the surface operator of full type \cite{Alday:2010vg}.
It is this second approach that we will take in this section . 
Since the surface operator of simple type and of full type coincide for $SU(2)$ theories,
we expect both descriptions give the same result. 
As we will see in the next subsection
the monopole expansion parameter $x$ comes from the $SU(2)$ spin variable that is
carried by the primary fields.


\subsection{Review of pure Yang-Mills case}

As argued by Alday-Tachikawa the $SU(2)$ instanton partition function with 
a surface operator is related to affine $\mathfrak{sl}(2)$ conformal
blocks \cite{Alday:2010vg}. 
For instance, the relation for the superconformal gauge theory with $N_f=4$ flavors takes the form
\beqa
\label{AT4f}
Z_{SU(2)}^{(S), N_f=4} 
=Z_{U(1)}^{(S)}
\langle V_{j_1}|V_{j_2}(1,1)\,\textbf{1}_{j}\,\mathcal{K}(x,z)V_{j_3}(x,z)|V_{j_4}\rangle,
\eeqa
where $V_{j}$ is the vertex operator with spin $j$.
$\textbf{1}_{j}$ is the projection operator on the Verma module spanned by  $V_{j}$,
and $\mathcal{K}(x,z)$ is the Alday-Tachikawa $K$-operator introduced in  \cite{Alday:2010vg}. 
The basic dictionary between gauge theory side and CFT side is
\begin{align}
j_1=-\frac{\epsilon_{+} +\mu_1-\mu_2}{2\epsilon_1},\quad
&j_2=-\frac{\epsilon_{+} +\epsilon_1+\mu_1+\mu_2}{2\epsilon_1},\nonumber\\
j_3=-\frac{\epsilon_{+} +\epsilon_1-\mu'_1-\mu'_2}{2\epsilon_1},\quad
&j_4=-\frac{\epsilon_{+} +\mu'_1-\mu'_2}{2\epsilon_1}, \label{ATdictionary} \\
j=-\frac{1}{2}+\frac{a_1}{\epsilon_1},\quad
&k=-2-\frac{\epsilon_2}{\epsilon_1}, \nonumber
\end{align}
where $a_1=-a_2=a$ as usual, and $k$ is the level of the affine algebra.
$\mu_{1,2}$ and $\mu'_{1,2}$ are the mass parameters for fundamental and anti-fundamental
matters.
The point in \cite{Alday:2010vg} is that
the conformal symmetry which controls the corresponding gauge theory
will change if we introduce a surface operator on the gauge theory side.
By introducing a surface operator,
the Virasoro algebra is replace by the (untwisted) affine $\mathfrak{sl}(2)$ algebra with 
the commutation relations;
\beqa
\left[ J_n^{0}, J_m^{0} \right] &=& \frac{k}{2} n \delta_{n+m,0}, \\
\left[ J_n^{0}, J_m^{\pm} \right] &=& \pm J_{n+m}^{\pm}, \\
\left[ J_n^{+}, J_m^{-} \right] &=& 2 J_{n+m}^{0} + k n \delta_{n+m,0}.
\eeqa
We can construct the Verma module of the affine $\mathfrak{sl}(2)$ algebra
as the highest weight representation of it.
The highest weight state $\vert j \rangle$ satisifies
\beq
J_0^{0} \vert j \rangle = j \vert j \rangle, \qquad
J_{1+n}^{-} \vert j \rangle = J_{1+n}^{0} \vert j \rangle = J_{n}^{+} \vert j \rangle = 0, \quad
(n \geq 0).
\eeq
Recall that the instanton partition function of an asymptotically free gauge theory
is believed to be equal to the norm of the corresponding Whittaker state.
As we will see soon, this correspondence will be extended to theories with surface operators.
For the pure $SU(2)$ Yang-Mills case
the Whittaker state is explicitly constructed in \cite{Kozcaz:2010yp}
(see also Braverman-Etingov \cite{Braverman:2004vv, Braverman:2004cr})
as follows;
\beq
J_{0}^{+} \vert x,z ; j \rangle = \sqrt{x} \vert x,z ; j \rangle, \qquad
J_{1}^{-} \vert x,z ; j \rangle = \sqrt{\frac{z}{x}} \vert x,z ; j \rangle,
\eeq
where $z$ and $x$ are identified with the parameters that count instanton and monopole number,
respectively, on the gauge theory side. 
In terms of the inverse of the Shapovalov matrix $Q_j( {\bf{n}}', {\bf{A}}' ; {\bf{n}}, {\bf{A}})$
of the Verma module on $\vert j \rangle$,
the Whittaker state is expanded as
\beq
\label{KPPWpure}
\vert x,z ; j \rangle = \sum_{n,p} \sum_{ {\bf{n}}, {\bf{A}}} x^{n/2} \left(\frac{z}{x}\right)^{p/2}
Q^{-1}_j (n , p  ; {\bf{n}}, {\bf{A}}) \vert {\bf{n}}, {\bf{A}} \rangle,
\eeq
where
\beq
\vert {\bf{n}}, {\bf{A}} \rangle := J_{-n_1}^{A_1} \cdots J_{-n_\ell}^{A_\ell} \vert  j \rangle
\eeq
is a basis of the Verma module\footnote{The ordering is $J_{-n}^{A}$ sits left of  $J_{-n'}^{A'}$
for $n > n'$ or $n=n', A < A'$.}
and
\beq
\vert n, p \rangle := (J_{-1}^{+})^{p} (J_{0}^{-})^{n} \vert  j \rangle \label{1st}
\eeq
is a special state with level $p$ and spin $p-n+j$. On the gauge theory side
$p$ counts the instanton number and $n-p$ counts the monopole number. 
The norm of the Whittaker state $\vert x,z ; j \rangle$ is computed as follows;
\beqa
&& \langle x,z ; j \vert x,z ; j \rangle \CR
&=& \sum_{n',p'} \sum_{ {\bf{n}'}, {\bf{A}'}}
\sum_{n,p} \sum_{ {\bf{n}}, {\bf{A}}} x^{(n+n')/2} \left(\frac{z}{x}\right)^{(p+p')/2}
Q^{-1}_j (n' , p'  ; {\bf{n}'}, {\bf{A}'})Q^{-1}_j (n , p  ; {\bf{n}}, {\bf{A}})
Q_j ({\bf{n}'}, {\bf{A}'} ; {\bf{n}}, {\bf{A}}) \CR
&=&  \sum_{n',p'} \sum_{n,p} x^{(n+n')/2} \left(\frac{z}{x}\right)^{(p+p')/2}
Q^{-1}_j (n , p  ; n' , p' )
= \sum_{n,p} x^{n} \left(\frac{z}{x}\right)^{p}
Q^{-1}_j (n , p  ; n , p). \label{expansion}
\eeqa
Thus we should compare $(n,p; n,p)$ component of $Q^{-1}_j $
with the partition function in the sector with instanton number $p$
and monopole number $n-p$. 
In \cite{Kozcaz:2010yp},
it is checked for special values of instanton and monopole numbers
that the irregular conformal block of the Whittaker state
gives the instanton partition function with a surface operator
\beqa
Z_{SU(2)}^{(S),N_f=0}
=
\langle x,z ; j \vert x,z ; j \rangle.
\eeqa
We will extend this relation to $SU(2)$ theory with fundamental flavor.
As a biproduct,
we find that the proposal of  \cite{Kozcaz:2010yp} follows precisely 
from the decoupling limit of Alday-Tachikawa conjecture.


\subsection{Decoupling limit of affine conformal block}
As we see in section 3 for $SU(3)$-$W_3$ case, 
the decoupling limit of fundamental hypermultiplets leads to the  Whittaker states
for asymptotically free gauge theories.
This idea also works for the gauge theories with a surface operator.
In this section, we construct the Whittaker states
by applying the decoupling limit to the correspondence 
(\ref{AT4f}) between the instanton partition function
with a surface operator
and the affine conformal block.
The four point conformal block in question is 
\begin{align}
\mathcal{B}&=
x^{-j +j_3+j_4}z^{\Delta -\Delta_3-\Delta_4}\nonumber\\
&\times
\sum_{{\bf{n}}, {\bf{A}}}
\langle\,j_1\,|
V_{j_2}(1,1)
|\,
{\bf{n}}, {\bf{A}};\,j
\,\rangle
Q^{-1}_j({{\bf{n}}, {\bf{A}};\,{\bf{n'}}, {\bf{A'}}})
\langle\,{\bf{n'}}, {\bf{A'}};\,j\,
|
\mathcal{K}(x,z)
V_{j_3}(x,z)
|\,j_4
\,\rangle.\nonumber
\end{align}
The vertex operator associated with the primary state satisfies
\begin{align}
[J_{-n}^{A},\,V_j(x,z)]
=z^{-n}D^{A}\,V_j(x,z),
\end{align}
where the representation of $\mathfrak{sl}(2)$ on the primary state is defined by
\begin{align}
D^{-}=\partial_x,\quad
D^0=-x\partial_x+j,\quad
D^+=2jx-x^2\partial_x.
\end{align}
We introduce the following three point spherical conformal blocks
for later consideration;
\begin{align}
&\rho(j_1;\,j_2;\,{\bf{n}}, {\bf{A}},\,j_3|\,x,z)
:=
\langle\,j_1\,|
V_{j_2}(x,z)
|\,
{\bf{n}}, {\bf{A}};\,j_3
\,\rangle,\\
&\rho^{\,{}_{\mathcal{K}}}({\bf{n}}, {\bf{A}},\,j_1;\,j_2;,\,j_3|\,x,z)
:=
\langle\,{\bf{n}}, {\bf{A}};\,j_1\,
|
\mathcal{K}(x,z)
V_{j_2}(x,z)
|\,j_3
\,\rangle.
\end{align}
These three point blocks are the basic building blocks 
of the above affine four point function $\mathcal{B}$.

We move on to explicit computation of the four point function
in asymptotically free gauge theory case.
In order to compute the decoupling limit,
let us derive the explicit expression of the three point function.
Since the action of the operator $J_0^0$ is given by
\begin{align}
\langle\,j_1\,|
[J_0^0,\,V_{j_2}(x,z)]&
|\,
{\bf{n}}, {\bf{A}};\,j_3
\,\rangle\nonumber\\
&=
\left(\langle\,j_1\,|J_0^0\right)
V_{j_2}(x,z)
|\,
{\bf{n}}, {\bf{A}};\,j_3
\,\rangle-
\langle\,j_1\,|
V_{j_2}(x,z)
J_0^0|\,
{\bf{n}}, {\bf{A}};\,j_3
\,\rangle,
\end{align}
the three point conformal block satisfies the differential equation
which determines the $x$-dependence 
\begin{align}
x\partial_x\,
\rho(j_1;\,j_2;\,{\bf{n}}, {\bf{A}},\,j_3|\,x,z)=
(-j_1+j_2+j_3+Q)\,\rho(j_1;\,j_2;\,{\bf{n}}, {\bf{A}},\,j_3|\,x,z).
\end{align}
The total charge is defined by $Q:=\sum_i A_i$.
This equation and the action of $L_0$ give the following $x$- and $z$-dependence of the block
\begin{align}
\rho(j_1;\,j_2;\,{\bf{n}}, {\bf{A}},\,j_3|\,x,z)=
x^{-j_1+j_2+j_3+Q}z^{\Delta_1-\Delta_2-\Delta_3-N}
\rho(j_1;\,j_2;\,{\bf{n}}, {\bf{A}},\,j_3).
\end{align}
Here we introduce
$
\rho(j_1;\,j_2;\,{\mathbf{n}}, {\mathbf{A}},\,j_3):= \rho(j_1;\,j_2;\,{\bf{n}}, {\bf{A}},\,j_3|\,1,1)
$.

The following is the derivation of the $\rho(j_1;\,j_2;\,{\bf{n}}, {\bf{A}},\,j_3|\,1,1)$ part of the three point block.
In order to determine an explicit form of the function,
we employ the recursive structure for the labels of the descendants
\begin{align}
\langle\,j_1\,|
V_{j_2}(x,z)
J_{-n}^{A}|\,
{\bf{n}}, {\bf{A}};\,j_3
\,\rangle&=-
\langle\,j_1\,|
[J_{-n}^{A},\,V_{j_2}(x,z)]
|\,
{\bf{n}}, {\bf{A}};\,j_3
\,\rangle+
\langle\,j_1\,|J_{-n}^{A}
V_{j_2}(x,z)
|\,
{\bf{n}}, {\bf{A}};\,j_3
\,\rangle\nonumber\\
&=-z^{-n}d^{A}\langle\,j_1\,|
V_{j_2}(x,z)
|\,
{\bf{n}}, {\bf{A}};\,j_3
\,\rangle,
\end{align}
where the coefficients are given by
\begin{align}
d^-=x^{-1}(-j_1+j_2+j_3+Q),\quad
d^0=j_1-j_3-Q,\quad
d^+=
x(j_1+j_2-j_3-Q).
\end{align}
When normalizing the three point function as 
$\rho(j_1;\,j_2;\,j_3)=1$, the recursive relation
leads to the following explicit expression of the three point conformal block:
\begin{align}
\rho(j_1;\,j_2;\,{\bf{n}}, {\bf{A}},\,j)
=(-j_1-j_2+j-c)_a\,
(-j_1+j-c)^b\,
(j_1-j_2-j)_c.
\end{align}
Here $a$, $b$ and $c$ are the number of $J^+$, $J^0$ and $J^-$ 
in the descendant $\vert {\bf{n}}, {\bf{A}}; j\rangle $ respectively\footnote{
Recall $[ * ]_n$ denotes the Pochhammer symbol.}.
Notice that we can rewrite the arguments by using gauge theory parameters:
\begin{align}
&j_1-j_2-j=1+\frac{\mu_2-a_1}{\epsilon_1}
,\nonumber\\
&j-j_1=
\frac{\epsilon_2+2a_1+\mu_1-\mu_2}{2\epsilon_1}
,\\
&j-j_1-j_2
=\frac{\epsilon_{+} +a_1+\mu_1}{\epsilon_1}. \nonumber
\end{align}


Let us proceed to the computation of the decoupling limit of a single hypermultiplet $\mu_1$ 
while keeping $\mu_2$ finite. 
In the decoupling limit $\mu_1\to\infty$,
the three point block behaves
\begin{align}
\rho(j_1;\,j_2;\,{\bf{n}}, {\bf{A}},\,j)\sim
\left(\frac{\mu_1}{\epsilon_1}\right)^{a+b}\frac{1}{2^b}
\left(1+\frac{\mu_2-a_1}{\epsilon_1}
\right)_c.
\end{align}
For a fixed charge $Q=a-c$ and level $N=\sum_{i,q}n_i^q$,
the dominant contribution in the limit comes from
the descendants with 
$n_i^+=n_i^0=1,n_i^-=0$.
These terms behave as $(\mu_1)^N$,
and then the conformal block $\mathcal{B}= \sum\rho\, Q^{-1} \rho^{\mathcal{K}}$ becomes
\begin{align}
\mathcal{B}\sim
&\sum_{N=0}^\infty
\sum_{a=0}^N
\sum_{Q=-\infty}^a
\left(\frac{\sqrt{z}\mu_1}{\epsilon_1}\right)^{N}
(\sqrt{x})^{-Q}
\frac{1}{2^{N-a}}
\left(1+\frac{\mu_2-a_1}{\epsilon_1}
\right)_{a-Q}
\nonumber\\
&\times
\sum_{\mathbf{n}, {\mathbf{A}}}
Q^{-1}_j(
{
(\underbrace {
\scriptstyle{
1,\cdots,1
}
}_{a};
\underbrace {
\scriptstyle{
1,\cdots,1
}
}_{N-a};
\underbrace{
\scriptstyle{0\cdots,0}}_{a-Q}),\,
(+,\cdots,+;0\cdots,0;-\cdots,-)
;\,\mathbf{n}, {\mathbf{A}}})\,
\rho^{{}_{\mathcal{K}}}(\mathbf{n}, {\mathbf{A}}).
\end{align}
The irregular conformal block takes the form $\sum\langle G_1,m\vert
\mathbf{n}, {\mathbf{A}}\rangle Q^{-1}({\mathbf{n}, {\mathbf{A}};
\mathbf{n'}, {\mathbf{A}'}})\rho^{{}_{\mathcal{K}}}(\mathbf{n'}, {\mathbf{A}'})$,
so that
the decoupling  limit $\mu\to \infty,\, \mu\sqrt{z}\to \sqrt{z}$ with fixed $x$ 
implies the following formula for the Whittaker-like state with one flavor
\begin{align}
|\,G_1,\,m\,\rangle
&=
\sum_{N=0}^\infty
\sum_{a=0}^N
\sum_{Q=-\infty}^a\sum_{\mathbf{n}, {\mathbf{A}}}
\left(\frac{\sqrt{z}}{\epsilon_1}\right)^{N}
(\sqrt{x})^{-Q}
\frac{1}{2^{N-a}}
\left(1+\frac{m-a_1}{\epsilon_1}
\right)_{a-Q}\nonumber\\
&\label{affineGW}
\times
Q^{-1}_j(
{
(\underbrace {
\scriptstyle{
1,\cdots,1
}
}_{a};
\underbrace {
\scriptstyle{
1,\cdots,1
}
}_{N-a};
\underbrace{
\scriptstyle{0\cdots,0}}_{a-Q}),\,
(+,\cdots,+;0\cdots,0;-\cdots,-)
;\,\mathbf{n}, {\mathbf{A}}})\,
|\, \mathbf{n}, {\mathbf{A}};\,j\rangle.
\end{align}
We derive the coherent condition for the state in the following.
Actually this state is not genuine Whittaker state,
and it turn out to be of generalized type.

Notice that by applying an additional decoupling limit $m\to\infty$,
the generalized Whittaker state $|G,m\rangle$ reproduces the proposal of \cite{Kozcaz:2010yp}
for pure $SU(2)$ Yang-Mills theory:
\begin{align}
|\,G_0\,\rangle
&=
\sum_{N=0}^\infty
\sum_{Q=-\infty}^N\sum_{\mathbf{n}, {\mathbf{A}}}
\left(\frac{\sqrt{z}}{\epsilon_1}\right)^{N}
(\sqrt{x})^{-Q}\nonumber\\
&\label{affineW}\times
Q^{-1}_j(
{
(\underbrace {
\scriptstyle{
1,\cdots,1
}
}_{N};
\underbrace{
\scriptstyle{0\cdots,0}}_{N-Q}),\,
(+,\cdots,+;0\cdots,0)
;\,\bf{n}, {\bf{A}}})\,
|\,\mathbf{n}, {\mathbf{A}};\,j\rangle.
\end{align}
This result is exactly equal to (\ref{KPPWpure}).


\subsection{Generalized Whittaker state from the decoupling}

From the computation of a decoupling limit in the last section, 
the Whittaker-like state with a matter hypermultiplet is identified with (\ref{affineGW}).
From the Shapovalov matrix of level zero, we find the level zero part of the state is
\beqa
\vert G_1, m \rangle_0  &=& \sum_{n=0}^\infty (\sqrt{x})^n  \left( 1 + \frac{m - a}{\ep} \right)_{n}
Q^{-1}_j{((J_0^{-})^{n} ; \mathbf{n}, \mathbf{A} )} \vert \mathbf{n}, \mathbf{A} ; j \rangle \\
&=& \sum_{n=0}^\infty \frac{(-\sqrt{x})^n}{n!} \frac{\left( 1 + \frac{m - a}{\ep} \right)_{n}}{\left( 1 - \frac{2a}{\ep} \right)_n}
(J_0^{-})^n \vert j \rangle, \label{level0}
\eeqa
which means
\beq
\langle j \vert (J_0^{+} )^n \vert G_1, m \rangle = (\sqrt{x})^n  \left( 1 + \frac{m - a}{\ep} \right)_{n}.
\eeq
Hence $ \vert G_1, m \rangle $ cannot be an eigenstate of $J_0^{+}$.

Since any state in the Verma module is expanded as 
\beq
\vert \Psi \rangle 
= \sum_{\mathbf{n}', \mathbf{A}'} \sum_{\mathbf{n}, \mathbf{A}} \vert \mathbf{n}', \mathbf{A}' ; j\rangle
Q^{-1}_j{( \mathbf{n}', \mathbf{A}' ; \mathbf{n}, \mathbf{A})}  \langle \mathbf{n} , \mathbf{A} ; j \vert \Psi \rangle, 
\eeq
at each level $N$, $\vert G_1, m \rangle$ is orthgonal to the subspace spanned by those states other than
$(J_{-1}^{+})^a (J_{-1}^{0})^{N-a} (J_0^{-})^{a+Q}  \vert j \rangle$. Namely
\beq
\label{orthogonal}
\langle  \mathbf{n} , \mathbf{A} ; j \vert G_1, m \rangle = 0,
\eeq
unless $\langle  \mathbf{n} , \mathbf{A} ; j \vert = \langle j \vert (J_0^+)^\ell (J_1^0)^m (J_1^{-})^n$ for 
which
\beq
 \langle j \vert (J_0^+)^\ell (J_1^0)^m (J_1^{-})^n \vert G_1,  m \rangle
 = \left( \frac{\sqrt{z}}{\ep} \right)^{m+n} (\sqrt{x})^{\ell - n} \frac{1}{2^m} \left( 1 + \frac{m -a}{\ep} \right)_\ell.
\eeq
Hence among the annihilation operators $J_{n}^{A}$, 
$J_{n}^{A} \vert G_1, m \rangle$ is non-vanishing only for $J_{n}^{A} =  J_0^+, J_1^0, J_1^{-}$.
To evaluate the action of these three operators on $\vert G_1, m \rangle$, let us look at
\beqa
\label{Jzeroplus}
 \langle j \vert (J_0^+)^\ell (J_1^0)^m (J_1^{-})^n  J_0^+ \vert G_1,  m \rangle &=& 
- 2n  \langle j \vert (J_0^+)^\ell (J_1^0)^{m+1} (J_1^{-})^{n-1}  \vert G_1,  m \rangle \nonumber \\
&&~~~~~~~
 +  \langle j \vert (J_0^+)^{\ell+1} (J_1^0)^m (J_1^{-})^n  \vert G_1,  m \rangle  \\
 &=& \sqrt{x} \left( \ell -n + 1 + \frac{m -a}{\ep}  \right)
   \langle j \vert (J_0^+)^{\ell} (J_1^0)^m (J_1^{-})^n  \vert G_1,  m \rangle, \nonumber \\
  \langle j \vert (J_0^+)^\ell (J_1^0)^m (J_1^{-})^n   J_1^0 \vert G_1,  m \rangle &=&   
  \langle j \vert (J_0^+)^\ell (J_1^0)^{m+1} (J_1^{-})^n   \vert G_1,  m \rangle \nonumber \\
  \label{Jonezero}
  &=&  \left( \frac{\sqrt{z}}{2\ep} \right)   \langle j \vert (J_0^+)^\ell (J_1^0)^m (J_1^{-})^n \vert G_1,  m \rangle, \\
   \langle j \vert (J_0^+)^\ell (J_1^0)^m (J_1^{-})^n   J_1^{-} \vert G_1,  m \rangle &=& 
     \langle j \vert (J_0^+)^\ell (J_1^0)^m (J_1^{-})^{n+1}  \vert G_1,  m \rangle \nonumber \\
     \label{Joneminus}
     &=&  \left( \frac{\sqrt{z}}{\ep\sqrt{x}} \right)   \langle j \vert (J_0^+)^\ell (J_1^0)^m (J_1^{-})^n \vert G_1,  m \rangle . 
\eeqa
By using (\ref{orthogonal}) we can easily see
$
\langle  \mathbf{n} , \mathbf{A} ; j\vert J_n^A \vert G_1, m \rangle = 0
$ unless $\langle  \mathbf{n} , \mathbf{A} ; j \vert = \langle j \vert (J_0^+)^\ell (J_1^0)^m (J_1^{-})^n
$
for $J_{n}^{A} =  J_0^+, J_1^0, J_1^{-}$.
Then, if $\langle j \vert (J_0^+)^\ell (J_1^0)^m (J_1^{-})^n J_n^A \vert G_1, m \rangle = C\langle j \vert (J_0^+)^\ell (J_1^0)^m (J_1^{-})^n \vert G_1, m \rangle$ is satisfied,
the Whittaker state is precisely the eigenstate of $J_{n}^{A}$:
\begin{align}
J_n^A \vert G_1, m \rangle&=
\sum
\vert \mathbf{n}' , \mathbf{A}' ; j \rangle
Q^{-1}_j({\mathbf{n}' , \mathbf{A}' ;\mathbf{n} , \mathbf{A} })
\langle  \mathbf{n} , \mathbf{A} ; j\vert J_n^A \vert G_1, m \rangle\nonumber\\
& = C
\sum
\vert \mathbf{n}' , \mathbf{A}' ; j \rangle
Q^{-1}_j({\mathbf{n}' , \mathbf{A}' ;\mathbf{n} , \mathbf{A} })
\langle  \mathbf{n} , \mathbf{A} ; j \vert G_1, m \rangle\nonumber\\
&=C \vert G_1, m \rangle.
\end{align}
This idea also works for the case when $C$ is a differential operator acting on $\vert G_1,m \rangle$.
Hence, the relations (\ref{Jzeroplus})-(\ref{Joneminus}) leads to the following relations;
\beqa
\label{J0+cond}
J_0^+ \vert G_1,  m \rangle &=& \left[  x \frac{\partial}{\partial(\sqrt{x})} 
+ \sqrt{x} \left( 1 + \frac{m -a}{\ep} \right) \right] \vert G_1,  m \rangle \label{J0+}, \\
 J_1^0 \vert G_1,  m \rangle &=&  \frac{\sqrt{z}}{2\ep} \vert G_1,  m \rangle,  \\
 J_1^{-} \vert G_1,  m \rangle &=&  \frac{\sqrt{z}}{\ep\sqrt{x}} \vert G_1,  m \rangle.
\eeqa
Compared with the pure Yang-Mills case\footnote{Note that the rescaling $\sqrt{z} \to \sqrt{z}/\ep$ makes
the parameter $z$ dimensionless.}, the crucial differences are a non-vanishing
eigenvalue of $J_1^0$ and the differential term in $J_0^+$. In fact they are not
unrelated,
because $J_1^0 \sim 1/2 [ J_0^+, J_1^-]$. A similar condition involving diffrential operators 
for the Virosoro CFT appeared recently in \cite{Bonelli:2011aa}.
Note that our Whittaker-like condition takes a very special form of generalization
involving the zero mode of $J^0$.
The Euler differential with the Coulomb moduli dependent term
in (\ref{J0+cond}) act as $J_0^0$
\beq
J_0^{0}  \vert G_1,  m \rangle 
=\left( j-\sqrt{x}\frac{\partial}{\partial \sqrt{x}}\right)
\vert G_1,  m \rangle,
\eeq
since the eigenvalue of $J_0^0$ on a descendant is $j+\sum A_i$.
We can therefore rewrite the first condition  (\ref{J0+cond}) into the following form
\beq
J_0^+ \vert G_1,  m \rangle
=\sqrt{x}\left(
\frac{1}{2}+ \frac{m}{\ep}
-J_0^0 
\right)\vert G_1,  m \rangle.
\eeq
In the right hand side, we have the zero-mode contribution $J_0^0$
as the case of the $W_3$ algebra,
and therefore $\vert G_1,  m \rangle$ is also of our generalized type introduced in section 3. 
The appearance of the zero-mode enables us to impose the eigenstate condition of $J_1^0$
in addition to $J_1^-\vert G_1,  m \rangle\propto\vert G_1,  m \rangle$:  
\beq
J_1^0 \vert G_1,  m \rangle
=\frac{1}{2}[J_0^+,J_1^-] \vert G_1,  m \rangle
=-\frac{\sqrt{x}}{2}[J_0^0,J_1^-] \vert G_1,  m \rangle
=\frac{\sqrt{x}}{2}J_1^-\vert G_1,  m \rangle.
\eeq
This is the very same phenomenon that we saw in section 3,
where $L_0$ makes it possible for the state
to be a simultaneous eigenstate of some annihilation operators.
These examples suggest that we have to introduce the generalized Whittaker states
of conformal algebras
for studying AGT-W correspondence in more generic set-up with matter fields.

We can check the level zero part of the Whittaker state \eqref{level0} satisfies \eqref{J0+} explicitly.
From
\beq
J_0^{+} (J_0^-)^n \vert j \rangle 
= n(2j-n+1) (J_0^-)^{n-1} \vert j \rangle 
= n \left( \frac{2a}{\ep} -n \right) \vert j \rangle
\eeq
we obtain
\beqa
J_0^{+} \vert G_1, m \rangle_0 &=& \sqrt{x} \sum_{n=1}^\infty \frac{(-\sqrt{x})^{n-1}}{(n-1)!} 
\frac{\left(1+ \frac{m -a}{\ep} \right)_n}{\left( 1 - \frac{2a}{\ep} \right)_{n-1}} (J_0^-)^{n-1} \vert j \rangle \CR
&=&
\sqrt{x} \left(1+ \frac{m -a}{\ep} \right) \sum_{n=1}^\infty \frac{(-\sqrt{x})^{n-1}}{(n-1)!} 
\frac{\left(1+ \frac{m -a}{\ep} \right)_{n-1}}{\left( 1 - \frac{2a}{\ep} \right)_{n-1}} (J_0^-)^{n-1} \vert j \rangle \CR
&&~~~+ \sqrt{x} \sum_{n=2}^\infty \frac{(-\sqrt{x})^{n-1}}{(n-2)!} 
\frac{\left(1+ \frac{m -a}{\ep} \right)_{n-1}}{\left( 1 - \frac{2a}{\ep} \right)_{n-1}} (J_0^-)^{n-1} \vert j \rangle.
\eeqa
On the other hand
\beqa
x \frac{\partial}{\partial \sqrt{x}} \vert G_1, m \rangle 
&=& -x \sum_{n=1}^\infty  \frac{(-\sqrt{x})^{n-1}}{(n-1)!} 
\frac{\left(1+ \frac{m -a}{\ep} \right)_{n}}{\left( 1 - \frac{2a}{\ep} \right)_{n}} (J_0^-)^{n} \vert j \rangle \CR
&=&
\sqrt{x} \sum_{n=1}^\infty  \frac{(-\sqrt{x})^{n}}{(n-1)!} 
\frac{\left(1+ \frac{m -a}{\ep} \right)_{n}}{\left( 1 - \frac{2a}{\ep} \right)_{n}} (J_0^-)^{n} \vert j \rangle.
\eeqa
This shows that  \eqref{level0} satisfies \eqref{J0+}.

Suppose we have the following states at level one\footnote{
See Appendix D for the notations of a basis of level one states. We have suppressed here 
the dependence of $Q_{ij}^{-1}$ on the monopole number $n$. };
\beqa
\vert \Psi_0 \rangle_1 &=&
\sum_{n=-1}^\infty a_n \sum_{k=1}^3 Q_{1k}^{-1} \vert k ; n, j \rangle
+ \sum_{n=0}^\infty \widetilde{a_n} \sum_{k=1}^3 Q_{2k}^{-1} \vert k ; n, j \rangle, \\
\vert \Psi_1 \rangle_1 &=&
\sum_{n=-1}^\infty b_n \sum_{k=1}^3 Q_{1k}^{-1} \vert k ; n, j \rangle
+ \sum_{n=0}^\infty \widetilde{b_n} \sum_{k=1}^3 Q_{2k}^{-1} \vert k ; n, j \rangle, 
\eeqa
then the scalar product of them is
\beq
\langle \Psi_0 \vert \Psi_1 \rangle_1 
= \sum_{n=-1}^\infty a_n b_n Q_{11}^{-1} + \sum_{n=0}^\infty (a_n \widetilde{b_n} + b_n \widetilde{a_n}) Q_{12}^{-1}
+ \sum_{n=0}^\infty \widetilde{a_n} \widetilde{b_n} Q_{22}^{-1}~.
\eeq
For pure Yang-Mills, we have $a_n = \left( \frac{\sqrt{z}}{\ep} \right) (\sqrt{x})^n, \widetilde{a_n}  =0$ and
for $N_f =1$ we take
\beq
b_n =  \left( \frac{\sqrt{z}}{\ep} \right) (\sqrt{x})^n \left( 1+ \frac{m -a}{\ep} \right)_{n+1}, \quad
\widetilde{b_n} =  \left( \frac{\sqrt{z}}{\ep} \right) \frac{(\sqrt{x})^n}{2} \left( 1+ \frac{m -a}{\ep} \right)_{n}. 
\eeq
We expect one instanton part of the partition function agrees with
\beqa
\langle G_0 \vert G_1, m \rangle_1 &=& \frac{z}{\ep^2} \left[ \sum_{n=-1}^\infty x^n \left( 1+ \frac{m -a}{\ep} \right)_{n+1} Q_{11}^{-1}
+ \frac{1}{2} \sum_{n=0}^\infty x^n \left( 1+ \frac{m -a}{\ep} \right)_{n} Q_{12}^{-1} \right] \CR
&=&  \frac{z}{x\ep^2} \frac{1}{k-2j} +  \frac{z}{\ep^2} \frac{k\left( 1 + \frac{m-a}{\ep} \right) -2j}{2j(k+2)(k-2j)} 
+ O(x).
\eeqa
Using the \lq\lq dictionary\rq\rq \eqref{ATdictionary} for $k$ and $j$,
we see
\beq
\langle G_0 \vert G_1, m \rangle_1 =  \frac{z}{-x\ep} \frac{1}{2a+\ep+\es} 
+  \frac{z}{-\ep} \frac{m(2\ep+\es) -a\es+ \ep^2 + \ep\es}{\es(2a-\ep)(2a+\ep+\es)} + O(x).
\eeq
After a redefinition
\beq
a \to -a - \frac{\es}{2}, \qquad m \to - m + \frac{\es}{2}, \label{redef}
\eeq
we find
\beq
\langle G_0 \vert G_1, m \rangle_1 =  \frac{z}{x\ep} \frac{1}{2a-\ep} 
-  \frac{z}{\ep} \frac{-m(2\ep+\es) +a\es+ (\ep+\es)^2}{\es(2a-\ep)(2a+\ep+\es)} + O(x).
\eeq
which agrees with the computation in \cite{Awata:2010bz}, where we keep $2M_2\equiv m$ finite
in the decoupling limit. We summarize the result of this decoupling limit in Appendix E. 

For general $n \geq 1$ the coefficient of $x^n$ is
\beq
\left( 1+ \frac{m -a}{\ep} \right)_{n}
\left[\left( n+1+ \frac{m -a}{\ep} \right)Q_{11}^{-1}(n) + \frac{Q_{12}^{-1}(n)}{2} \right], \label{general}
\eeq
where the components $Q_{ij}^{-1}(n)$ of the inverse of the level one Shapovalov matrix are
given in Appendix D.  After the redefinition \eqref{redef} we can check an exact agreement of \eqref{general} 
with the result of the same decoupling limit in Appendix E. 


\section{Conclusion and discussion}
For ${\mathcal N}=2$ superconformal gauge theories in four dimensions,
the AGT-W conjecture claims the equivalence between the instanton partition functions and
the conformal blocks of corresponding CFT.
Pure Yang-Mills theory also has a natural formulation in terms of CFT:
the partition functions are given by the norms of certain coherent states of the
conformal algebra of the CFT.
Such a special state in the Verma module is called the Whittaker state.
Between these two extreme cases,
there exists a series of the asymptotically free gauge theories with hypermultiplets.
So far, the CFT formulation of these series was studied only for $SU(2)$ gauge theory
without a surface operator.
We therefore tried to make the generic feature of the 4d/2d correspondence clear
by studying asymptotically free gauge theories with fundamental flavors.

Our strategy we adopted in this paper is as follows;
we start with assuming the AGT-W relation for superconformal cases $N_f=2N_c$.
Then the decoupling limit leads to irregular conformal blocks which is expected to be equal to
the instanton partition functions of asymptotically free gauge theories.
We regard this conformal block as the norm of a certain state in the Verma module.
For pure Yang-Mills theories, this is just the Gaiotto-Whittaker state
and our results provide a generalization to the case with fundamental matter. 
We have worked out the conditions that our state satisfies and 
introduced the notion of the generalized Whittaker states. Namely
for the $SU(3)$ gauge theory with $N_f=2$,
we found a little non-familiar relation:
\beqa
(W_1-wi\Lambda L_0) |\,G_2,\,m_1, m_2\rangle =
\frac{3 wi \Lambda}{2}\left(2m_1m_1-Q(m_1+m_2)+Q^2\right)
  |\,G_2,\,m_1, m_2\rangle.\nonumber
\eeqa
Similarly, for $SU(2)$ gauge theory with a surface operator, $N_f=1$ flavor
is encoded into the state $ \vert G_1,  m \rangle$ that is characterized by
the conditions with the zero mode $J_0^0$ of the $\mathfrak{sl}(2)$ current algebra.
These two examples suggest the generality
of the existence of the generalized Whittaker states 
in CFT description for gauge theories with flavors. 

In section 5, we saw that the scalar product of the generalized Whittaker states
 $\langle G_0\vert G_1\rangle$
 reproduces the instanton partition function for $SU(2)$ theory with $N_f=1$ flavor.
The irregular conformal block $\langle G_0\vert G_0\rangle$ also gives that for pure Yang-Mills.
Note that an explicit combinatorial formula for the same partition function
$Z_{SU(2)}^{(S), N_f=0}$ of the Yang-Mills theory was proposed in \cite{Taki:2010bj}.
Though this combinatorial formula is essentially based on another description (\ref{simpleS0}) of surface operators
which realizes them as the degenerates field of Virasoro CFT,
we can see in the lower order of the instanton expansion 
this approach implies the same answer as the affine Whittaker state gives.
It would be interesting to study such an equivalence between affine Whittaker states
and Virasoro Gaiotto states with $\Phi_{1,2}$ for the theory with flavors and extend the combinatorial formula
to these cases.

We may be able to define the generalized Whittaker states for more generic chiral algebra of CFT
such as $W$-algebra.
As argued in section 4 the M-theoretic construction of $SU(N)$ theory suggests that
the Whittaker-like state for $N_f=N-1$ flavors would be of the generalized type.
Let us consider a simultaneous eigenstate
for $W_N$ algebra: $L_1|G\rangle=\ell_1|G\rangle$, $L_2|G\rangle=\ell_2|G\rangle$.
Since the spin-$s$ current satisfies $[L_n,W^{(s)}_m]=((s-1)n-m)W^{(s)}_{n+m}$,
we can impose the following conditions:
\beq
W^{(s)}_1|G\rangle=(w_1+w_0L_0)|G\rangle\quad
\Longrightarrow\quad
W^{(s)}_2|G\rangle=\frac{\ell_1w_0}{s-2}|G\rangle,\quad
W^{(s)}_3|G\rangle=\frac{2\ell_2w_0}{2s-3}|G\rangle.
\eeq
Actually, all conditions must be consistent with the commutation relations
of modes of currents, which are nonlinear and very complicated.
It will be very cumbersome to write down the full series of the consistent conditions,
but we expect it is doable in principle.
They will provide the irregular conformal blocks 
which coincide with instanton partition functions for more generic $SU(N)$ gauge theories.
We also expect such generalized Whittaker states exist
for BCDEFG gauge groups \cite{Keller:2011ek} and 
it is an interesting challenge to carry out these extensions. 
%



\section*{Acknowledgments}

We would like to thank A.~Marshakov, K.~Maruyoshi, A.~Mironov, A.~Morozov
 and Y.~Tachikawa for enlightening discussions. 
The work is supported in part by Grant-in-Aid for Scientific Research
[No.22224001], JSPS Bilateral Joint Projects (JSPS-RFBR collaboration) 
  (H.K.) and JSPS Grant-in-Aid for Creative Scientific Research No.19GS0219 (M.T.)
from MEXT, Japan. 

\newpage

\section*{Appendix A : Building block of the Nekrasov partition function}

\renewcommand{\theequation}{A.\arabic{equation}}\setcounter{equation}{0}
\renewcommand{\thesubsection}{A.\arabic{subsection}}\setcounter{subsection}{0}

The instanton partition function introduced by Nekrasov \cite{Nekrasov:2002qd}
is an integral of certain volume form over the instanton moduli space.
In the strict sense this is defiend as an equivariant integral,
and we can evaluate it with the localization formula.
The resulting generic formula for $SU(N)^p$ linear quiver gauge theory takes the form\footnote{
$N_{f,i}$ is the flavor number for $i$-th gauge group.
For conformal gauge group $2N=N_{f,i}$,
we introduce the gauge coupling constant 
in compensation for vanishing dynamical scale as $\Lambda_i^{2N-N_{f,i}}\to q_i=e^{2\pi i\tau_i}$.}
\beqa
&&Z_{SU(N)^p,\textrm{lin.quiver}}
=\sum_{\vec{Y}_{1,2,\cdots,p}}\prod_{i=1}^p \Lambda_i^{(2N-N_{f,i})\vert\vec{Y}_i\vert}
\prod_{\tilde{f}=1}^{\tilde{F}} z_{\mathrm{antifun}}(\vec{a}_1, \vec{Y}_1: \tilde{m}_{\tilde{f}} ) \,
z_{\mathrm{vec}}(\vec{a}_1, \vec{Y}_1 ) \nonumber\\
&&\qquad\qquad\times
z_{\mathrm{bif}} (\vec{a}_1, \vec{Y}_1 ; \vec{a}_2 ,\vec{Y}_2 ; \mu_1 )\,
z_{\mathrm{vec}}(\vec{a}_2, \vec{Y}_2 )\,
z_{\mathrm{bif}} (\vec{a}_2, \vec{Y}_2 ; \vec{a}_3 ,\vec{Y}_3 ; \mu_2 )\,
z_{\mathrm{vec}}(\vec{a}_3, \vec{Y}_3 ) \nonumber\\
&&\qquad\quad\quad\,\,\,\,\,\times\cdots
z_{\mathrm{bif}} (\vec{a}_{p-1}, \vec{Y}_{p-1} ; \vec{a}_p ,\vec{Y}_p ; \mu_{p-1} )\,
z_{\mathrm{vec}}(\vec{a}_p, \vec{Y}_p )\,
\prod_{f=1}^Fz_{\mathrm{fun}}(\vec{a}_p, \vec{Y}_p: m_f ).
\eeqa
The weight factor $z_*$,
which is a combinatorial rational function of gauge theory parameters, represents the contribution 
from the vector or hyper multiplet labeled by $*$. 
Since $\Lambda_i$ is the dynamical scale for $i$-th gauge group,
$|\vec{Y}_i|$ is the instanton number of the gauge group factor.
The $k_i$-instanton partition function is therefore defined by summing the Young diagrams
with fixing the number of boxes as $|\vec{Y}_i|=k_i$.
We use the following weight factor 
in the localization formula of the Nekrasov function;

The contribution of a bifundamental matter multiplet is
\beqa
z_{\mathrm{bif}} (\vec{a}, \vec{Y} ; \vec{b} ,\vec{W} ; m )
&=& \prod_{\alpha, \beta = 1}^N
 \prod_{(i,j) \in Y_\alpha} 
(a_\alpha- b_{\beta} - m + \ep( - W^t_{\beta, j} +i) + \es (Y_{\alpha, i} - j+1)) \CR
&&~~\times \prod_{(i,j) \in W_\beta}
(a_\alpha- b_{\beta} - m + \ep( Y^t_{\alpha, j} -i +1) + \es (-W_{\beta, i} + j)),
\eeqa
The contributions of an adjoint matter and a vector matter multiplet are
related to $z_{\mathrm{bif}} (\vec{a}, \vec{Y} ; \vec{b} ,\vec{W} ; m )$ by
\beq
z_{\mathrm{adj}}(\vec{a}, \vec{Y} ; m )
= z_{\mathrm{bif}} (\vec{a}, \vec{Y} ; \vec{a} ,\vec{Y} ; m),
\quad
z_{\mathrm{vec}}(\vec{a}, \vec{Y} ) 
= \frac{1}{z_{\mathrm{adj}} (\vec{a}, \vec{Y} ;  0 )}.
\eeq
Finally a fundamental matter and an anti-fundamental matter contribute
\beq
z_{\mathrm{fun}}(\vec{a}, \vec{Y}: m ) 
= \prod_{\alpha=1}^N \prod_{(i,j) \in Y_\alpha} (a_\alpha -m + \ep i + \es j),
\quad
z_{\mathrm{anti}}(\vec{a}, \vec{Y}: m ) = z_{\mathrm{fun}}(\vec{a}, \vec{Y}: -m + \ep +\es ).
\eeq


\section*{Appendix B : Explicit 1-instanton Check of Wyllard conjecture}

\renewcommand{\theequation}{B.\arabic{equation}}\setcounter{equation}{0}
\renewcommand{\thesubsection}{B.\arabic{subsection}}\setcounter{subsection}{0}

In section 3, we apply the decoupling limit to the derivation of the Whittaler states.
The decoupling limit is also useful to simplify the check of Wyllard's proposal
because the original set-up is too complicated to verify by explicit computation.

At one instanton level, 
the Wylalrd conjecture suggests the relation between the 1-instanton
partition function $Z_{k=1}$ for the $SU(3)$ superconformal SQCD
and the level-1 ${W}_3$ conformal block $\mathcal{B}_{1}$
\begin{align}
\mathcal{B}^{N_f=6}_{\,1} =Z_{SU(3),k=1}^{N_f=6}+\nu,
\end{align}
for the following $U(1)$ factor \cite{Mironov:2009by}
\begin{align}
Z_{U(1)}=(1-q)^{-\nu},\quad
\nu
=\frac{1}{4}(\sqrt{3}Q-2\alpha_1)(\sqrt{3}Q+2\alpha_3).
\end{align}
In the 2-flavor limit 
\begin{align}
\mu_1\mu_4\mu_5\mu_6q
\rightarrow
\mu_1\mu_4\mu_5\Lambda_{N_f=5}
\rightarrow
\mu_1\mu_4\Lambda_{N_f=4}^2
\rightarrow
\mu_1\Lambda_{N_f=3}^3
\rightarrow
\Lambda_{N_f=2}^4,
\end{align}
the $U(1)$ factor disappears,
and we expect the following relation for the 1-instanton sector of SQCD with two anti-fundamental flavors
\begin{align}
\lim_{2\textrm{-flavor}}\mathcal{B}^{N_f=6}_{\,1}=Z_{SU(3),k=1}^{N_f=2}.
\end{align}
This relation must hold if the Wyllard conjecture is true.
Let us prove it.

\subsection{1-instanton Nekrasov partition function}
The one-instanton partition function of the $SU(3)$ gauge theory with two anti-fundamentals is
\begin{align}
Z_{SU(3),k=1}^{N_f=2}\,(\epsilon_{1,2})
=
\Lambda_{N_f=2}^4
\sum_{i=1}^3
\frac{(a_i+\mu_1)(a_i+\mu_2)}{\prod_{j\neq i}a_{ij}(a_{ij}+\epsilon_+)},
\end{align}
where $a_{ij}=a_i-a_j$ for $i,j=1,2,3$.
For the self-dual $\Omega$-background $\epsilon_+(=Q)=0$,
the partitin function becomes very simple
\begin{align}
\label{2-flav1-instNek}
&Z_{SU(3),k=1}^{N_f=2}\,(\epsilon_+=0)\nonumber\\
&=
\frac{\Lambda_{N_f=2}^4}{(a_{12}a_{23}a_{31})^2}
\Big[ 
(a_1+a_2)^4+a_1^4+a_2^4
-9(\mu_2+\mu_3)a_1a_2a_3
+6\mu_2\mu_3(a_1^2+a_1a_2+a_2^2)
\Big].
\end{align}
We can easily check that this result recovered from the CFT side only by hand.

\subsection{Level-1 irregular conformal block}
The level-1 conformal block is
\begin{align}
\mathcal{B}^{N_f=6}_{\,1} &=
q\big(\,\Delta+\Delta_1-\Delta_2,\,
w+2w_1-w_2+\frac{3w_1}{2\Delta_1}(\Delta-\Delta_1-\Delta_2)
\,\big)\,
(Q_{\Delta({\vec{\alpha}})}^{(1)})^{-1} \nonumber \\
&~~~~\times \left(\begin{array}{c}
\Delta+\Delta_3-\Delta_4
 \\
 w+w_3+w_4-\frac{3w_3}{2\Delta_3}(\Delta+\Delta_3-\Delta_4)
 \end{array}\right).
\end{align}
Let us take the 3-flavor limit $\mu_{4,5,6}\to \infty$ first.
By using the result in \cite{Taki:2009zd},
we get
\begin{align}
\mathcal{B}_{\,1}^{N_f=3} 
&=
\Lambda_{N_f=3}^3\big(\,\Delta+\Delta_1-\Delta_2,\,
w+2w_1-w_2+\frac{3w_1}{2\Delta_1}(\Delta-\Delta_1-\Delta_2)
\,\big)\, \nonumber \\
&~~~~ \times (Q^{(1)}_{\Delta({\vec{\alpha}})})^{-1}
\left(\begin{array}{c}
0
 \\
\frac{3\sqrt{3}}{\sqrt{4-15Q^2}}
 \end{array}\right).
\end{align}
Then by applying the 2-flavor limit 
$\lim \mu_1\Lambda_{N_f=3}^3=-\lim\sqrt{3}A\Lambda_{N_f=3}^3=\Lambda_{N_f=2}^4$,
we get
\begin{align}
&\mathcal{B}_{\,1}^{N_f=2}\nonumber\\
&=
\frac{-3}{\sqrt{4-15Q^2}}
\Lambda_{N_f=2}^4\frac{1}{\textrm{det}Q_{\Delta({\vec{\alpha}})}^{(1)}} \nonumber \\
&~~~\times
\left(\,2C,\,
\frac{9\sqrt{\kappa}}{4}(2\mu_2\mu_3-Q(\mu_2+\mu_3)+Q^2
+\frac{2}{3}(a_1^2+a_1a_2+a_2^2-Q^2)\,\right)
\,
\left(\begin{array}{c}
-3w
 \\
2\Delta
 \end{array}\right)\nonumber\\
 &=
\Lambda_{N_f=2}^4 \frac{1}
{(a_{12}+Q)(a_{12}-Q)(a_{23}+Q)(a_{23}-Q)(a_{31}+Q)(a_{31}-Q)}\nonumber\\
&\rule{0pt}{5ex}\times
\Big[
(6\mu_2\mu_3-3Q(\mu_2+\mu_3)+3Q^2
+2(a_1^2+a_1a_2+a_2^2-Q^2))(a_1^2+a_1a_2+a_2^2-Q^2)\nonumber\\
&\rule{0pt}{5ex}\qquad+
9(\mu_2+\mu_3-Q)a_1a_2(a_1+a_2)
\Big].
\end{align}
Of course this result is consistent with the generic formula we get in section.3.
By setting $Q=0$ we recover the gauge theory result (\ref{2-flav1-instNek}).
We can also check the generic case $Q\neq0$.



\section*{Appendix C: Conditions for $\vert G_2,m_1, m_2 \rangle$}

\renewcommand{\theequation}{C.\arabic{equation}}\setcounter{equation}{0}
\renewcommand{\thesubsection}{C.\arabic{subsection}}\setcounter{subsection}{0}

The following arguments are quite parallel to those in section 5.3.
The point is that  the formula \eqref{Gstate2} in section 3 implies the following non-vanishing inner product
\beqa
\label{naiseki}
& & \langle \Delta(\vec{\alpha}) \vert W_1^t W_2^q W_3^p
L_1^s L_2^r \vert G_2, m_1, m_2 \rangle \CR
& &~~~ =  (i\Lambda)^{3p+2q+2r+s+t} 
\frac{2^{p} (\sqrt{3})^{-3p+q-2r+t}}{(\sqrt{4-15 Q^2})^{p+q+t}} 
 (q_L(m_1, m_2))^{s+q} \left( \frac{3}{2} q_W(m_1, m_2) \right)_{t},
\eeqa
where
\beqa
q_L(m_1, m_2)  &:=& Q - m_1 - m_2,
\\
q_W(m_1, m_2)  &:=& 2 m_1 m_2 - Q(m_1 + m_2) + Q^2 + \frac{2}{3} (a_1^2 + a_1 a_2 + a_2^2 - Q^2),
\eeqa
and $\vert G_2, m_1, m_2 \rangle$ is orthogonal to other states in the Verma module.
By this orthogonality and the commutation relation of $W_3$ algebra, we can compute
the result of the insertion of $W_2$ as follows;
\beqa
&& \langle \Delta(\vec{\alpha}) \vert W_1^t W_2^q W_3^p
L_1^s L_2^r W_2 \vert G_2, m_1, m_2 \rangle 
=
\langle \Delta(\vec{\alpha}) \vert W_1^t W_2^{q+1} W_3^{p}
L_1^{s}  L_2^{r}  \vert G_2, m_1, m_2 \rangle \CR
&& \rule{0pt}{5ex} \hspace{10mm}=  \frac{\sqrt{3}(i\Lambda)^2}{\sqrt{4 -15Q^2}} q_L (m_1, m_2) 
\langle \Delta(\vec{\alpha}) \vert W_1^t W_2^{q} W_3^{p}
L_1^{s}  L_2^{r}  \vert G_2, m_1, m_2 \rangle,
\eeqa
where the commutation relation $[L_1, W_2] = (2\cdot 1-2) W_{1+2} =0$ is crucial.
We can read the eigenvalue of $W_2$ on the state $\vert G_2, m_1, m_2 \rangle$ from
this relation. Similarly we compute $L_{1,2}$ and $W_{3}$ insertions to obtain
their eigenvalues. The computation of $W_1$ insertion is special in the sense 
that the contributions from the commutation relations survive;
\beqa
\label{W1insertion}
& &\langle \Delta(\vec{\alpha}) \vert W_1^t W_2^q W_3^p
L_1^s L_2^r  W_1 \vert G_2, m_1, m_2 \rangle \CR
& & =  \frac{\sqrt{3}i\Lambda}{\sqrt{4- 15 Q^2}} \left( \frac{3}{2} q_W(m_1, m_2) + t \right) 
\langle \Delta(\vec{\alpha}) \vert W_1^t W_2^q W_3^p
L_1^s L_2^r \vert G_2, m_1, m_2 \rangle+ \cdots
\eeqa
where the additional contributions $\cdots$ come from the commutation relations
\beqa
&& [L_2, W_1] = 3 W_3, \qquad [L_1, W_1] = W_2, \qquad [L_1, [L_1, W_1]] =0,  \\
&& [W_3, W_1] \sim \frac{2 \cdot 9}{4-15Q^2} (L_2)^2, \qquad \left[W_2, W_1\right] \sim \frac{9}{4-15Q^2} 2 L_1 L_2. 
\eeqa
Note that the $t$-dependent term in \eqref{W1insertion} comes from the Pochhammer product in \eqref{naiseki}. 
Using \eqref{naiseki}, we can check the contribution from the commutation relation $[ X_n, W_1]$ for $X=L,W$
is always proportional to 
$\langle \Delta(\vec{\alpha}) \vert W_1^t W_2^q W_3^p
L_1^s L_2^r  \vert G_2, m_1, m_2 \rangle$ and
the coefficient is  $\frac{n\sqrt{3}i\Lambda}{\sqrt{4 - 15 Q^2}}$ times the number of $X_n$ between
$\vert \Delta(\vec{\alpha}) \rangle$ and $\vert G_2, m_1, m_2 \rangle$. For example,
the contribution from $[W_2, W_1]$ is 
\beqa
%
&& \frac{18q}{4-15Q^2} \langle \Delta(\vec{\alpha}) \vert W_1^{t} W_2^{q-1} W_3^{p}
L_1^{s+1}  L_2^{r+1}  \vert G_2, m_1, m_2 \rangle \CR
&=& \frac{2q\sqrt{3}i\Lambda}{\sqrt{4 - 15 Q^2}} 
\langle \Delta(\vec{\alpha}) \vert W_1^t W_2^q W_3^p
L_1^s L_2^r  \vert G_2, m_1, m_2 \rangle.
\eeqa
Hence, combined with the $t$ dependent term in \eqref{W1insertion}, 
the sum of the additional contributions is neatly expressed by the action of the Euler derivative
\beq
\frac{\sqrt{3}i\Lambda}{\sqrt{4 - 15 Q^2}} \Lambda \frac{\partial}{\partial \Lambda}, 
\eeq
that counts the level of the state in the Verma module. 
We further note that the last term of $q_W(m_1, m_2)$ is proportional to  the eigenvalue of $L_0$ 
on the primary state. Together with this part the Euler derivative gives
the action of the Virasoro zero mode $L_0$.


\section*{Appendix  D: Level one Shapovalov matrix of $SU(2)$ current algebra}

\renewcommand{\theequation}{D.\arabic{equation}}\setcounter{equation}{0}
\renewcommand{\thesubsection}{D.\arabic{subsection}}\setcounter{subsection}{0}

We summarize the data of the Shapovalov matrix at level one which corresponds to
one instanton sector with monopole number $\mathfrak{m} = n-1 \geq -1$.
At level one, we have three states with spin $1-n+j$,
\beq
\vert 1 \rangle = J_{-1}^{+} (J_0^{-})^n \vert j \rangle, \quad
\vert 2 \rangle =  J_{-1}^{0} (J_0^{-})^{n-1} \vert j \rangle, \quad
\vert 3 \rangle =  J_{-1}^{-} (J_0^{-})^{n-2} \vert j \rangle. 
\eeq
Using commutation relations, we obtain the Shapovalov matrix. 
When $n \geq 2$ the Shapovalov matrix is 
\beq
Q(n)= 
\left(\begin{array}{ccc}
(k -2j +2n) M(n)  &  M(n) & 0 \\
M(n) & \frac{k}{2} M(n-1) & - M(n-1) \\
0 & - M(n-1) & (k+2j-2(n-2)) M(n-2) 
\end{array}\right),
\eeq
with
\beq
M(n) := n ! (-1)^n (-2j)_n,
\eeq
where $(X)_n := X(X+1) \cdots (X+n-1)$. The determinant of $Q(n)$ factorizes as follows;
\beq
\mathrm{det}~Q(n) = \frac{1}{2} (k+2) (2j+k+2)(k-2j) M(n-2) M(n-1) M(n).
\eeq
When $n=0,1$ the Shapovalov matrix is reduced to one by one, or two by two matrix.
\beq
Q(0)= 
\left(\begin{array}{c}
k -2j \\
\end{array}\right),
\qquad
Q(1)= 
\left(\begin{array}{cc}
 2j(k -2j +2) &  2j\\
 2j& \frac{k}{2} \\
\end{array}\right).
\eeq
The inverse of the Shapovalov matrix for $n=0,1$ is simple. 
The components of the inverse of the Shapovalov matrix for $n \geq 2$ are
\beqa
Q(n)^{-1}_{11} &=&  \frac{2n^2 -2(k +3 +2j) n+2j(k+2)+ (k+2)^2}{(k+2)(2j+2+k)(k-2j)M(n)},
\\
Q(n)^{-1}_{12} &=&  \frac{-2(k+2j-2n+4)}{(k+2)(2j+2+k)(k-2j)M(n-1)},
\\
Q(n)^{-1}_{13} &=&  \frac{-2}{(k+2)(2j+k+2)(k-2j)M(n-2)},
\\
Q(n)^{-1}_{22} &=&  \frac{-2(-k+2j -2n)(k+2j-2n+4)}{(k+2)(2j+k+2)(k-2j)M(n-1)},
\\
Q(n)^{-1}_{23} &=&  \frac{-2(-k+2j-2n)}{(k+2)(2j+k+2)(k-2j)M(n-2)},
\\
Q(n)^{-1}_{33} &=& \frac{2 n^2 - 2(1 - k +2j)n -2kj + k^2}{(k+2)(2j+k+2)(k-2j)M(n-2)}.
\eeqa


\section*{Appendix E: Partition function of $SU(2)$ theory with a surface operator}

\renewcommand{\theequation}{E.\arabic{equation}}\setcounter{equation}{0}
\renewcommand{\thesubsection}{E.\arabic{subsection}}\setcounter{subsection}{0}

We here quote the result presented in section 8 of \cite{Awata:2010bz},
where we computed the partition function of superconformal ($N_f=4$) $SU(2)$ theory
with a surface operator at one instanton and arbitrary monopole number by localization.
As is common in $SU(2)$ theories the fixed points of the torus action on the moduli space
are labeled by a pair of Young diagrams (partitions) $\vec{\lambda} = (\lambda_1, \lambda_2)$.
We identify the instanton number $k$ and the monopole number $\mathfrak{m}$ 
at each fixed point by:
\beq
k = k_1, \qquad \mathfrak{m} = k_2 - k_1,
\eeq
where $k_1$ and $k_2$ are given by
\beq
k_1(\vec\lam) = \sum_{n \geq 1} \lam_{1, 2n-1} + \sum_{n \geq 1} \lam_{2,2n}, 
\quad k_2(\vec\lam) = \sum_{n \geq 1} \lam_{1, 2n} + \sum_{n \geq 1} \lam_{2,2n-1}.
\eeq
Thus in one instanton sector with $k_1=1,\quad k_2=\mathfrak{m}+1$, 
there are four choices for $\vec{\lambda}$ as follows:
\begin{eqnarray}
&&(A)\quad \vec{\lambda}_{\mathfrak{m}A}=((1),(\mathfrak{m}+1)),\quad \mathfrak{m}\ge -1,  \qquad
(B)\quad \vec{\lambda}_{\mathfrak{m}B}=(\emptyset,(\mathfrak{m}+1,1)),\quad \mathfrak{m}\ge 0,
\nonumber \\
&&(C)\quad \vec{\lambda}_{\mathfrak{m}C}=(\emptyset,(\mathfrak{m},1,1)), \quad \mathfrak{m}\ge 1, \qquad
(D)\quad \vec{\lambda}_{\mathfrak{m}D}=((1,1),(\mathfrak{m})), \quad \mathfrak{m}\ge 0.
\label{fixed}
\end{eqnarray}
Taking a decoupling limit
\begin{eqnarray}
M_1,M_3,M_4\to \infty,
\quad \widetilde{z} =4 M_1M_3x, \quad
\widetilde{x} = -2 M_4y,
\end{eqnarray}
in the result of section 8 of \cite{Awata:2010bz},
one obtains the following contributions to the partition function for $N_f=1$ theory at one instanton
\begin{eqnarray}
&&
Z_1^{(A)}(a,M,\epsilon_1,\epsilon_2; \widetilde{z}, \widetilde{x})
\nonumber \\
&&\quad
=\sum_{\mathfrak{m}=-1}^{\infty} \widetilde{z}~\widetilde{x}^{\mathfrak{m}+1}
\frac{- \prod_{k=1}^{\mathfrak{m}+1}(a-2M+k\epsilon_1+\epsilon_2)}
{(2a+\mathfrak{m}\epsilon_1)\epsilon_1^{\mathfrak{m}+2}(\mathfrak{m}+1)!\prod_{k=0}^{\mathfrak{m}}(2a+k\epsilon_1+\epsilon_2)},
\nonumber \\
&&
Z_1^{(B)}(a,M,\epsilon_1,\epsilon_2; \widetilde{z}, \widetilde{x})
\nonumber \\
&&\quad
=\sum_{\mathfrak{m}=0}^{\infty}\widetilde{z}~\widetilde{x}^{\mathfrak{m}+1}
\frac{\prod_{k=1}^{\mathfrak{m}+1}(a-2M+k\epsilon_1+\epsilon_2)}
{(\mathfrak{m}\epsilon_1-\epsilon_2)\epsilon_1^{\mathfrak{m}+1}
\mathfrak{m}!\prod_{k=0}^{\mathfrak{m}+1}(2a+k\epsilon_1+\epsilon_2)},
\nonumber\\
&&
Z_1^{(C)}(a,M,\epsilon_1,\epsilon_2; \widetilde{z},\widetilde{x})
\nonumber \\
&&\quad=\sum_{\mathfrak{m}=1}^{\infty} \widetilde{z}~\widetilde{x}^{\mathfrak{m}+1}
\frac{ (a-2M+\epsilon_1+2\epsilon_2)
\prod_{k=1}^{\mathfrak{m}}(a-2M+k\epsilon_1+\epsilon_2)}
{(2a+\epsilon_1+2\epsilon_2)
(-\mathfrak{m} \epsilon_1+\epsilon_2)\epsilon_2
\epsilon_1^{\mathfrak{m}}(\mathfrak{m}-1)!\prod_{k=0}^{\mathfrak{m}}(2a+k\epsilon_1+\epsilon_2)},
\nonumber\\
&&
Z_1^{(D)}(a,M,\epsilon_1,\epsilon_2; \widetilde{z}, \widetilde{x})
\nonumber \\
&&\quad=\sum_{\mathfrak{m}=0}^{\infty}\widetilde{z}~\widetilde{x}^{\mathfrak{m}+1}
\frac{(a+2M-\epsilon_1-\epsilon_2)
\prod_{k=1}^{\mathfrak{m}}(a-2M+k\epsilon_1+\epsilon_2)}
{(2a-\epsilon_1)(2a+\mathfrak{m}\epsilon_1)\epsilon_2
\epsilon_1^{\mathfrak{m}+1}\mathfrak{m}!\prod_{k=0}^{\mathfrak{m}-1}(2a+k\epsilon_1+\epsilon_2)}.
\label{1-instanton_Nf=1}
\end{eqnarray}
The one instanton partition function $Z_{SU(2), k=1}^{(S), N_f=1}$ is the sum of the above contributions
and a formal power series in $x$. 
Note that the number of the fixed points is reduced for lower monopole numbers $\mathfrak{m} = -1, 0$. 



\end{document}